\documentclass[twocolumn,superscriptaddress,aps,prb,longbibliography]{revtex4-2}
\usepackage[utf8]{inputenc}
\usepackage[T1]{fontenc}
\usepackage{newtxtext,newtxmath}
\usepackage{graphicx}
\usepackage{hyperref}
\usepackage{bm}
\usepackage{microtype}
\usepackage{physics}
\usepackage{xcolor}
\usepackage[caption=false]{subfig}
\usepackage{relsize}
\newcommand{\bigast}{\mathlarger{\ast}}

\begin{document}
\title{A proposal for realizing Majorana fermions without external magnetic field in strongly correlated nanowires}

\author{Kaushal K. Kesharpu}
\email{kesharpu@theor.jinr.ru}

\author{Evgenii A. Kochetov}
\affiliation{Bogoliubov Laboratory of Theoretical Physics, Joint Institute for Nuclear Research, Dubna, Moscow Region 141980, Russia}

\author{Alvaro Ferraz}
\affiliation{International Institute of Physics - UFRN, Department of Experimental and Theoretical Physics - UFRN, Natal 59078-970, Brazil}

\date{\today}

\begin{abstract}

We show that one dimensional (1D) topological superconductivity can be placed in the context of phenomena associated with strongly correlated electron systems. Here we propose a system consisting of a one-dimensional chain of strongly correlated fermions placed on a superconducting (SC) substrate that exhibits a spin-singlet extended $s$-wave pairing. Strong electron correlation is shown to transform an extended $s$-wave SC into a topological SC that hosts Majorana fermions. In contrast to the approaches based on mean-field treatments, no Zeeman or exchange magnetic field is needed to produce such an effect.

\end{abstract}

\maketitle

\section{Introduction}
The topological superconductivity with Majorana zero modes (MZM) is of fundamental scientific importance due to its proposed application in braiding-based quantum computing \cite{sarma_2015_MajoranaZero_NpjQuantumInf}. The first theoretical proposal for the realization of the one dimensional (1D) MZM was involved by placing a 1D quantum wire with spinless electrons on a $p$-wave superconductor \cite{kitaev_2001_UnpairedMajorana_Phys-Usp}. However, from the material science point of view this conceptually simple model is hard to realize because a spinless electron does not exist in nature by default, and $p$-wave superconductors are, at best, rare to find. Later several proposals were put forward to eliminate these difficulties by ingeniously combining the proximity induced $s$-wave superconductivity, the Rashba spin-orbit coupling (RSOC), and the breaking of time reversal (TR) symmetry \cite{alicea_2012_NewDirections_RepProgPhys}. In these proposed systems, RSOC was needed for the spin-momentum locking \cite{manchon_2015_NewPerspectives_NatureMater}, and TR symmetry breaking was needed to create, in effect, the spinless electrons. Using aforementioned ideas, broadly three types of platforms have been engineered till now, by placing \cite{flensberg_2021_EngineeredPlatforms_NatRevMater}: (i) topological insulators on a superconductor \cite{hasan_2010_ColloquiumTopological_RevModPhys,qi_2011_TopologicalInsulators_RevModPhys}, (ii) semiconductors with strong spin-orbit coupling on a superconductor \cite{lutchyn_2018_MajoranaZero_NatRevMater}, or (iii) chains of magnetic atoms on superconductor \cite{nadj-perge_2013_ProposalRealizing_PhysRevB,nadj-perge_2014_ObservationMajorana_Science}.

In heterostructures, where the topological insulator is placed over the superconductor, the 1D helical edge states provide the needed spin-momentum locking \cite{qi_2011_TopologicalInsulators_RevModPhys}. The 1D edge states occur due to RSOC. The TR symmetry in these heterostructures can be broken, either by applying a magnetic field \cite{xu_2015_ExperimentalDetection_PhysRevLett,lutchyn_2010_MajoranaFermions_PhysRevLett,sau_2010_RobustnessMajorana_PhysRevB}, or by either doping with magnetic atoms \cite{yu_2022_TopologicalSuperconductivity_}, or by placing the magnetic materials in the vicinity of the topological insulators \cite{fu_2008_SuperconductingProximity_PhysRevLett,he_2014_SelectiveEqualSpin_PhysRevLett}. In the case of a semiconductor on superconductor heterostructures usually a semiconductor wire with strong RSOC and magnetic field are used to satisfy the aforementioned requirements to generate MZM \cite{mourik_2012_SignaturesMajorana_Science,lutchyn_2018_MajoranaZero_NatRevMater}. For devices with a magnetic ad-atom chain on a superconductor, the Yu-Shiba-Rusinov (YSR) states induce spin polarized subgap in the parent superconductors \cite{nadj-perge_2014_ObservationMajorana_Science,nadj-perge_2013_ProposalRealizing_PhysRevB}. If the magnetic ad-atoms are placed very close to each other the YSR states hybridize, and the hopping between neighboring ad-atoms broadens the particle-hole symmetric YSR pair states into bands. Once the band formed by the positive and negative energies overlap at the center of the superconducting gap, the $p$-wave pairing correlations will reopen the gap and turn the ad-atom chain into a 1D topological superconductor with MZM. The $p$-wave correlation can be induced either by RSOC in the parent superconductor, or by the helical spin states on the magnetic ad-atom chain \cite{choy_2011_MajoranaFermions_PhysRevB}.

From  the experimental side, claims of MZM in topological insulators (Bi/Nb \cite{jack_2019_ObservationMajorana_Science}, Bi$_{2}$Te$_{3}$ \cite{chen_2009_ExperimentalRealization_Science,xu_2015_ExperimentalDetection_PhysRevLett}, Bi$_{2}$Se$_{3}$ \cite{xia_2009_ObservationLargegap_NaturePhys,chen_2012_RobustnessTopological_ProcNatlAcadSci}, and Bi$_{2}$Te$_{3-x}$Se$_{x}$ \cite{yu_2022_TopologicalSuperconductivity_}) have been made. For semiconductors on superconductor platforms usually InAs and InSb on superconductors have been extensively used in the experiments \cite{lutchyn_2018_MajoranaZero_NatRevMater}. Regarding the magnetic ad-atoms on superconductors, the chain of Fe atoms on either Pb \cite{nadj-perge_2014_ObservationMajorana_Science}, Re \cite{kim_2018_TailoringMajorana_SciAdv}, or Ta \cite{kamlapure_2018_EngineeringSpin_NatCommun} substrates have shown signs of the MZM. Despite these experimental successes, several concerns still remain which have to do mainly with the requirements of the strong RSOC and the external magnetic field. The former limits the candidate materials which can be placed over the superconductor, while the latter limits the possible superconducting substrates. Besides the crystal symmetry consideration greatly shrink the possible superconductors supporting topological phases \cite{ono_2020_RefinedSymmetry_SciAdv}. The external magnetic field is the principal hindrance, because most experiments use $s$-wave superconductors for which their superconductivity might be destroyed in a strong enough magnetic field. Hence, naturally the question arises, can we get rid of the constraint imposed by the magnetic field? It is the goal of this article to suggest a different route to realizing the MZM without making use of an external magnetic field.

In a strong magnetic field the spins of the electrons are aligned along the direction of the applied field. The Pauli constraint prohibits the occupation of more than one fermion in a single quantum state. Hence no more than one electron with the same spin can occupy a single site. Physically, it is the same as having spinless fermions. Therefore one way of realizing the spinless fermions is through the application of a magnetic field. Another way to produce such an effect is to rely on the presence of a large on-site Coulomb repulsion. In such a strong correlation regime each lattice site becomes at most singly occupied, and the charge and spin degrees of freedom separate from each other. The resulting fermion is effectively spinless and this is precisely what we need. The prototype model for such a behavior is the $U \to \infty$ Hubbard model \cite{arovas_2022_HubbardModel_AnnuRevCondensMatterPhys,imada_1998_MetalinsulatorTransitions_RevModPhys,roy_2019_MottInsulators_}. The main reason why such an approach was not used before, was probably due to inadequacy of the mean-field methods to treat strong correlation. In this paper we show that the strong electron correlation can indeed drive the 1D system of lattice electrons into the topological phase as the Kitaev's toy model suggests \cite{kitaev_2001_UnpairedMajorana_Phys-Usp}. This consideration is based on the representation of the underlying dynamics in terms of Hubbard operators that encode strong electron correlation effects. The basic requirement of no local double occupancy is taken into account rigorously prior to any mean-field treatments. To this end the $su(2|1)$ coherent-state action is employed, where the $su(2|1)$ superalgebra incorporates the full set of Hubbard operators \cite{wiegmann_1988_SuperconductivityStrongly_PhysRevLett}. We propose a device, in which a strongly correlated nanowire is placed over an extended $s$-wave superconductor. Usually iron based superconductors fulfill this requirement \cite{mazin_2011_SymmetryAnalysis_PhysRevB,hirschfeld_2011_GapSymmetry_RepProgPhys,stewart_2011_SuperconductivityIron_RevModPhys}. The details of experimental realization is discussed in Sec. \ref{sec:prop-exper-heter}.

It should be mentioned that an analogous idea was presented before \cite{zhang_2013_TimeReversalInvariantTopological_PhysRevLett}. However, there instead of a strongly correlated nanowire, a semiconducting nanowire with strong RSOC on an extended $s$-wave superconductor was proposed in its place. Hence, the problem of using magnetic field still persisted (although critical magnetic field of iron-based superconductors are quite high $\sim 50 \: T$ \cite{hosono_2018_RecentAdvances_MaterialsToday}). In another analogous work it was predicted that a semiconducting nanowire with $C_{n}$ ($n$>4) rotational symmetry and non-zero electronic angular momentum ($l \neq 0$) develops MZM when Coulomb repulsion is high enough \cite{li_2019_CoulombinteractioninducedMajorana_PhysRevB}. In this case magnetic field, RSOC, and even superconductivity is not needed either. Another interesting idea is to use profiled topological insulators \cite{thalmeier_2021_SurfaceStep_PhysRevB}; it does not need the RSOC or magnetic field.

A similar system as our was considered in Ref. \cite{aksenov_2020_StrongCoulomb_PhysRevB} numerically. Using DMRG they showed that mean-field approximation is not valid in strong correlation regime $(U/t \leq 2)$, and also MZM can be produced in this regime. Analogous work in Ref. \cite{stoudenmire_2011_InteractionEffects_PhysRevB} showed that, due to strong correlation the parameter space of chemical potential magnetic field for generation of the MZM increases. In fact, they even predicted that the MZM can be created only due to the strong correlation --- without magnetic field. Our article differs from all the above works, in the sense, we explicitly solve the corresponding Hamiltonian analytically using $su(2)$ coherent state representation of the Hubbard $X^{pq}$ operators, and show the presence of the MZM. In the same framework we discuss the effect of RSOC on the system.

This article is structured as follows. In Sec. \ref{sec:model} we found the effective Hamiltonian for the physical system shown in Fig. \ref{fig:schematic}. In Sec. \ref{sec:emerg-topol-effects} we derived the expression of the Hamiltonian when the spiral spin field is present in the wire. In Sec. \ref{sec:absence-rsoc-alpha=0} and \ref{sec:presence-rsoc-alpha} we discussed the case when RSOC in the wire is both absent and present respectively. In Sec. \ref{sec:internal-energy} we calculated the thermodynamic free energy of the nanowire as a function of system parameters, and discuss its consequences. In Sec. \ref{sec:prop-exper-heter} we propose experiments to observe the Majorana fermions, and the schematics for device design. In Sec. \ref{sec:conclusion} we summarized our results.

\section{Model}
\label{sec:model}
\begin{figure}
  \centering
  \includegraphics[width=0.4\textwidth]{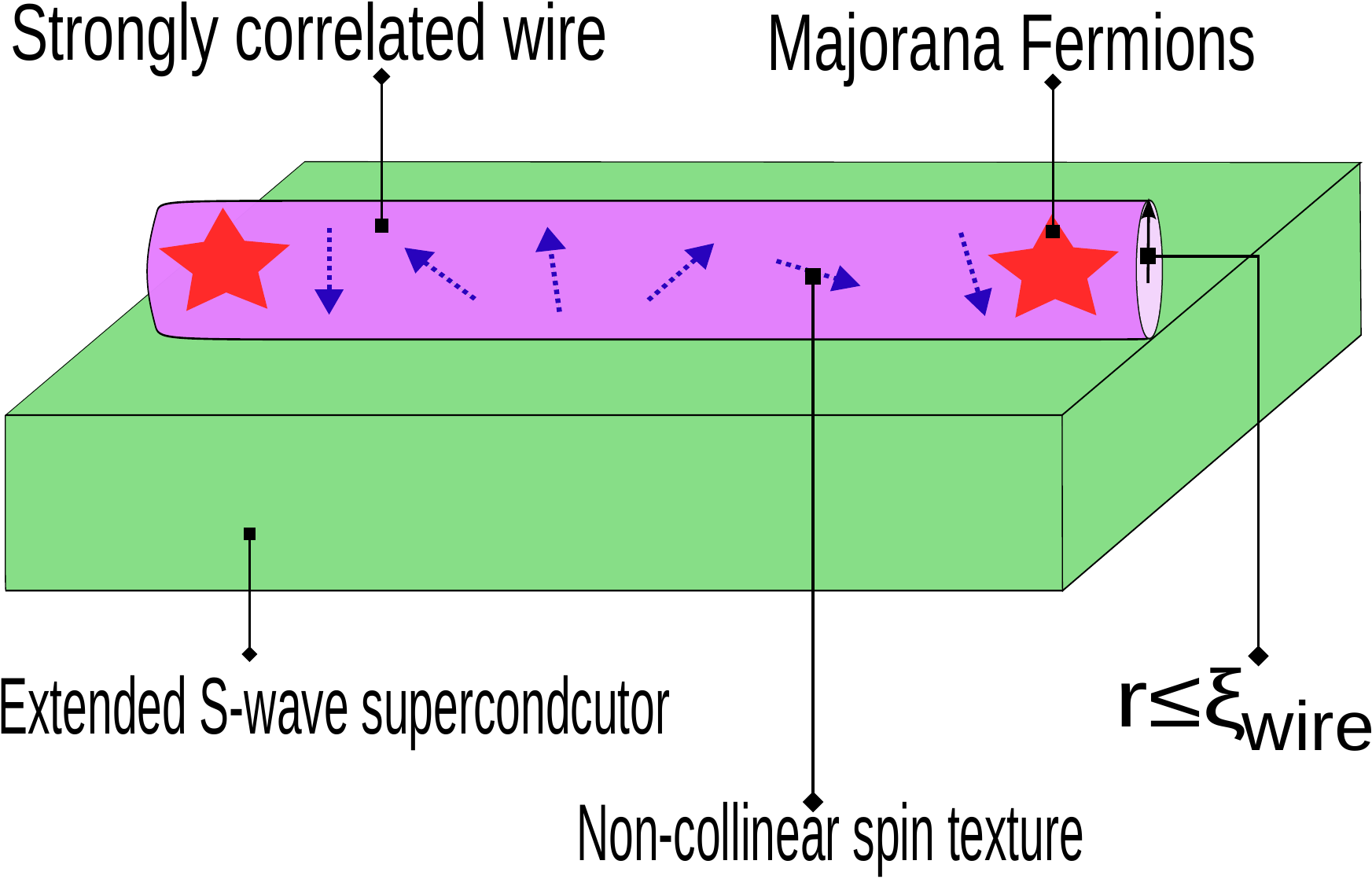}
  \caption{A schematic of the system under consideration. A strongly correlated 1D nanowire is placed over a superconductor with an extended $s$-wave order parameter. Superconductivity is induced in the nanowire due to the proximity effect. The radius of the wire ($r$) is assumed to be smaller than the Cooper pair coherence length in the wire ($\xi_{wire}$): $r \lessapprox \xi_{wire}$. The Majorana fermions appear at the end of the nanowire (stars) if a non-colinear spin texture is present in the wire.}
  \label{fig:schematic}
\end{figure}
The schematics of the system under consideration is shown in Fig. \ref{fig:schematic}. Here, a strongly correlated 1D nanowire with RSOC is placed on a superconducting substrate. Due to the proximity of the wire to the superconductor we assume tunneling of Cooper pair into the wire (tunneling of single electron is neglected due to the reasons discussed further). Under these assumptions, the total Hamiltonian of the wire will be:
\begin{equation}
  \label{eq:Ham-1d-wire}
  \begin{aligned}
    &H &&= H_{1D} + H_{\alpha} + H_{\Gamma};\\
    \text{where,}\\
    &H_{1D} &&= -t \sum\limits_{i, \sigma} X_{i}^{\sigma 0}X_{i+1}^{0 \sigma}  + \text{H.c} + \mu \sum\limits_{i} X_{i}^{00};\\
    &H_{\alpha} &&= - \imath \alpha \sum\limits_{i} X_{i}^{\sigma' 0}X_{i+1}^{0 \sigma} - X_{i}^{\sigma 0}X_{i+1}^{0 \sigma'} + \text{H.c};\\
    &H_{\Gamma} &&= \Gamma \sum\limits_{i, \sigma} X_{i}^{\sigma 0} d_{i \sigma} + \text{H.c}.
  \end{aligned}
\end{equation}
Here, $t$ is the electron hopping factor, and $\mu$ is the chemical potential. $\sigma$ is the electron spin which can have only two quantum states $\uparrow$ and $\downarrow$. $X_{i}^{pq} = \bra{p}\ket{q}$ is the Hubbard operator, which represent the transition from the $\ket{q}$ state to $\ket{p}$ state at the $i$-th site. In the strong correlation regime, i.e. when Coloumb repulsion between electrons is infinitely large, there only three states are allowed at each atomic sites: $\ket{\uparrow}$, $\ket{\downarrow}$, or $\ket{0}$. For example, $X_{i}^{\uparrow 0}$ represent the transition from empty state to the up-spin state at the $i$-th site. In terms of usual electron creation ($c_{i\sigma}^{\dagger}$), annihilation ($c_{i\sigma}$), and number ($n_{i\sigma} = c_{i \sigma}^{\dagger} c_{i \sigma}$) operators, the Hubbard operators are represented as:
  \begin{equation}
    \label{eq:fn:Hubbard-opr}
    \begin{aligned}
      &X_{i}^{0\sigma} = c_{i\sigma} (1-n_{i \sigma'}),
      \quad X_{i}^{\sigma 0} =(1-n_{i \sigma'})c_{i\sigma}^{\dagger},\\
      &X_{i}^{0 0} = X_{i}^{0\sigma} X_{i}^{\sigma 0} = c_{i\sigma} (1-n_{i \sigma'}) (1-n_{i \sigma'})c_{i\sigma}^{\dagger}.
    \end{aligned}
  \end{equation}
Above definition of Hubbard operators shows that the no double occupancy condition is satisfied. $X_{i}^{00}$ gives the occupation number of the $i$-th site. It should be remembered that the electron is created only if the $i$-th site was empty initially. In the total Hamiltonian $H$, the $H_{1D}$ part contains the kinetic energy and chemical potential of the wire.

$H_{\alpha}$ represents the RSOC in the wire. $\alpha$ is the Rashba parameter which depends on the crystal electric field generated due to the breaking of the inversion symmetry \cite{bihlmayer_2022_RashbalikePhysics_NatRevPhys}; quantitatively, $\alpha \sim 1$ eV \cite{manchon_2015_NewPerspectives_NatureMater}. $H_{\Gamma}$ corresponds to the tunneling of electrons from the superconducting substrate to the wire with amplitude $\Gamma$.  $d_{i\sigma}$ is the electron annihilation operator with spin $\sigma$ at $i$-th site of the superconducting substrate. $X_{i}^{\sigma 0}d_{i\sigma}$ means the annihilation of the electron in the substrate and creation of the same in the wire. It is important to note that the Hubbard operators $X_i^{\sigma 0}$ transform themselves in the lowest (fundamental) representation of the $SU(2)$ group \cite{scheunert_1977_GradedLie_JMathPhys}, so that Eq. (\ref{eq:Ham-1d-wire}) represents a rotationally invariant interaction.

As the dimension and physical size of the wire is smaller than the substrate, it is natural to assume that the Fermi level of the superconductor greatly exceeds that of the wire, $k_{\text{F}}^{\text{SC}} \gg k_{\text{F}}^{\text{wire}}$. The large difference in Fermi levels allows the hybridization between them to be controlled mainly by the superconducting energy gap ($\Delta_{\text{SC}}$) and $\Gamma$. For small $\Gamma$, i.e. for $\Gamma k_{\text{F}}^{\text{SC}} \ll \Delta_{\text{SC}}$, we can treat $H_{\Gamma}$ peturbatively \cite{alicea_2012_NewDirections_RepProgPhys}. The first order perturbation terms can be neglected, due to the suppression of the single electron tunneling \footnote{
Due to induced superconducting gap $\Delta$ at the Fermi level of the wire the energy penalty for exciting a single electronic state from the top of the valence band to the bottom of the conduction band in the wire is too high. It is due to our assumption $k_{\text{F}}^{\text{SC}} \gg k_{\text{F}}^{\text{wire}}$, which makes sure that $E_{F}^{\text{wire}} \sim \Delta$.
}. The second order perturbation generates an effective Cooper pair hopping \cite{hekking_1994_SubgapConductivity_PhysRevB,hekking_1993_CoulombBlockade_PhysRevLett}. Due to the presence of strong correlation one can discard the single site spin-singlet $s$-wave pairing $\left( \Delta_{i,i}^{\uparrow \downarrow} \right)$ contribution \footnote{ A pure spin-singlet $s$-wave pairing $\Delta^{\uparrow\downarrow}_{i,i}$ does not contribute into $\delta H$ since
it is multiplied by a corresponding bilinear form of the Hubbard operators that disappear at the coinciding lattice sites, e.g., $X^{\sigma 0}_i X^{\sigma' 0}_i =0.$ This is a direct manifestation of strong electron correlation that prohibits on-site double electron occupancy due to infinitely strong on-site Coulomb repulsion. In this case, the proximity-induced pairing is only possible provided the bulk superconductor develops either the spin-triplet $p$-wave pairing or an extended $s$-wave one. A pure $s$-wave pairing is totally excluded.
}. Therefore, on neighboring sites two types of spin pairing are possible: (i) spin-singlet extended $s$-wave pairing and (ii) spin-triplet $p$-wave pairing. The corresponding second order perturbed Hamiltonian is:
\begin{equation}
  \label{eq:petrub-ham}
  \begin{aligned}
    \delta H_{\Gamma} \propto &  \sum\limits_{i} \Delta_{i,i+1}^{\sigma \sigma'} \: \left( X_{i}^{\sigma 0} X_{i+1}^{\sigma' 0} -  X_{i}^{\sigma' 0} X_{i+1}^{\sigma 0}  \right) \\
                     & + \sum\limits_{i} D_{i,i+1}^{\sigma \sigma'} \: \left( X_{i}^{\sigma 0} X_{i+1}^{\sigma' 0} +  X_{i}^{\sigma' 0} X_{i+1}^{\sigma 0}  \right)\\
    & \quad + \sum\limits_{i,\sigma} D_{i,i+1}^{\sigma \sigma} \: X_{i}^{\sigma 0} X_{i+1}^{\sigma 0} + \text{H.c.}
  \end{aligned}
\end{equation}
Here, $D_{i,i+1}^{\sigma \sigma}$, $D_{i,i+1}^{\sigma \sigma'}$ are the $p$-wave order parameters, and $\Delta_{i,i+1}^{\sigma \sigma'}$ is the extended $s$-wave order parameter. In low energy limit we can approximate the order parameters through ground state expectation values:
\begin{equation}
  \label{eq:expect-val-order-param}
  \begin{aligned}
    D_{i,i+1}^{\sigma \sigma} &= \left\langle d_{i, \sigma} d_{i+1, \sigma}\right\rangle\\
    D_{i,i+1}^{\sigma \sigma'} &= \left\langle d_{i, \sigma} d_{i+1, \sigma'} + d_{i, \sigma'} d_{i+1, \sigma} \right\rangle\\
    \Delta_{i,i+1}^{\sigma \sigma'} &= \left\langle d_{i, \sigma} d_{i+1, \sigma'} - d_{i, \sigma'} d_{i+1, \sigma} \right\rangle.
  \end{aligned}
\end{equation}
In Eq. (\ref{eq:expect-val-order-param}) the spin-triplet $p$-wave order parameters are symmetric in spin $\left(D_{i,i+1}^{\sigma \sigma'} = D_{i,i+1}^{\sigma' \sigma}\right)$, but anti-symmetric in space $\left(D_{i,i+1}^{\sigma \sigma'} = -D_{i+1,i}^{\sigma \sigma'}\right)$. On the other hand the extended $s$-wave order parameters are anti-symmetric in spin $\left(\Delta_{i,i+1}^{\sigma \sigma'} = -\Delta_{i,i+1}^{\sigma' \sigma}\right)$, but symmetric in space $\left(\Delta_{i,i+1}^{\sigma \sigma'} = \Delta_{i+1,i}^{\sigma \sigma'}\right)$. Usually $p$-wave pairing in superconductor is quite rare and unstable, on the other hand, the extended $s$-wave pairing is predicted in high $T_{c}$ iron based superconductors \cite{stewart_2017_UnconventionalSuperconductivity_AdvPhys,stewart_2011_SuperconductivityIron_RevModPhys}. Hence, from now on we assume that, the substrate develops only extended $s$-wave pairing. Considering above discussion we can approximate:
\begin{equation}
  \label{eq:ham-sc}
  \delta H_{\Gamma} = \Delta \sum\limits_{i}  X_{i}^{\sigma 0} X_{i+1}^{\sigma' 0} -  X_{i}^{\sigma' 0} X_{i+1}^{\sigma 0} + \text{H.c.}
\end{equation}
Here $\Delta$ is the induced superconducting gap in the wire. Its value can be estimated through $\Delta \approx \Gamma^{2}/k_{\text{F}}^{\text{SC}}\Delta_{\text{SC}}$. We have assumed spatially independent order parameter $\Delta_{i,i+1}^{\sigma \sigma'} \equiv \Delta_{\text{SC}}$.
Combining Eq. (\ref{eq:Ham-1d-wire}) and Eq. (\ref{eq:ham-sc}) we write the effective Hamiltonian of the wire:
\begin{equation}
  \label{eq:ham-1d-wire}
  \begin{aligned}
    H_{\text{eff}} =& -t \sum\limits_{i, \sigma} X_{i}^{\sigma 0}X_{i+1}^{0 \sigma} \\
                    &\quad- \imath \alpha \sum\limits_{i} X_{i}^{\sigma' 0}X_{i+1}^{0 \sigma} - X_{i}^{\sigma 0}X_{i+1}^{0 \sigma'}\\
                    &\qquad + \Delta \sum\limits_{i}  X_{i}^{\sigma 0} X_{i+1}^{\sigma' 0} -  X_{i}^{\sigma' 0} X_{i+1}^{\sigma 0} + \text{H.c.}\\
                    &\qquad \quad + \mu \sum\limits_{i} X_{i}^{00}.
  \end{aligned}
\end{equation}

To solve the Hamiltonian, Eq. (\ref{eq:ham-1d-wire}), the $su(2|1)$ path integral representation of the partition function will be used. This is possible due to the existence of an injective mapping between Hubbard operators $X^{\sigma \sigma'}$ and their $su(2|1)$ coherent-state symbols \cite{ferraz_2011_EffectiveAction_NuclearPhysicsB,kochetov_1995_SUCoherent_JournalOfMathematicalPhysics,wiegmann_1988_SuperconductivityStrongly_PhysRevLett} (also known as Berezin symbols \cite{berezin_1972_COVARIANTCONTRAVARIANT_MathUSSRIzv}):
\begin{equation}
  \label{eq:coher-symb}
  X^{\sigma \sigma'} (z, \xi) \equiv \bra{z, \xi} X^{\sigma \sigma'} \ket{z, \xi}.
\end{equation}
Here $z$ and $\xi$ are the complex even and odd Grassmann parameters respectively. The state $\ket{z, \xi}$ is defined as:
  \begin{equation}
    \begin{aligned}
      \ket{z,\xi}& \equiv \left( 1 + \bar{z} z +\bar{\xi} \xi \right)^{-1/2} \exp \left( z X^{\downarrow \uparrow} + \xi X^{0 \uparrow} \right) \ket{\uparrow}\\
      & = \left( 1 + \bar{z} z +\bar{\xi} \xi \right)^{-1/2} \left(  \ket{\uparrow} + z \ket{\downarrow} + \xi \ket{0} \right).
    \end{aligned}
  \end{equation}
Above representation was found by acting with the lowering superspin operators $X^{\downarrow 0}$ and $X^{\downarrow \uparrow}$ on the highest weight state $\ket{\uparrow}$. For a complete exposition on this subject please see Ref. \cite{ferraz_2011_EffectiveAction_NuclearPhysicsB,ferraz_2022_FractionalizationStrongly_PhysRevB}. Physically, $z$ represents the spinfull bosonic fields, and $\xi$ represents the spinless charged fermionic fields \cite{kesharpu_2023_TopologicalHall_PhysRevB,ferraz_2022_FractionalizationStrongly_PhysRevB}. In this new representation and using the change of variables
\begin{equation*}
  \label{eq:chng-variable}
  \xi \to \frac{\xi}{\sqrt{1+\left| z \right|^{2}}}, \quad z \to z,
\end{equation*}
the effective Hamiltonian Eq. (\ref{eq:ham-1d-wire}) will be:
\begin{equation}
  \label{eq:effective-ham-cov}
  \begin{aligned}
    H_{\text{eff}}(z, \xi) =& -t \sum\limits_{i} \xi_{i} \bar{\xi}_{i+1} a_{i,i+1} + \imath \alpha \sum\limits_{i} \xi_{i} \bar{\xi}_{i+1} \alpha^{\ast}_{i,i+1} \\
    &\qquad - \Delta\sum\limits_{i}\xi_{i}\xi_{i+1} \Delta^{\ast}_{i,i+1} + \text{H.c.} + \mu \sum\limits_{i} \bar{\xi}_{i} \xi_{i},
  \end{aligned}
\end{equation}
where,
\begin{subequations}
  \begin{equation}
    \label{eq:def-aij}
    a_{i,i+1} \equiv \frac{1 + \bar{z}_{i} z_{i+1}}{\sqrt{\left( 1 + \left| z_{i} \right|^{2}\right) \left( 1 + \left| z_{i+1} \right|^{2}\right)}},
  \end{equation}
  \begin{equation}
    \label{eq:def-alpha-ij}
    \alpha^{\ast}_{i,i+1} \equiv \left[ \frac{z_{i} - \bar{z}_{i+1}}{\sqrt{\left( 1 + \left| z_{i} \right|^{2}\right) \left( 1 + \left| z_{i+1} \right|^{2}\right)}} \right]^{\bigast},
  \end{equation}
  \begin{equation}
    \label{eq:def-delta-ij}
    \Delta^{\ast}_{i,i+1} \equiv \left[   \frac{z_{i} - z_{i+1}}{\sqrt{\left( 1 + \left| z_{i} \right|^{2}\right) \left( 1 + \left| z_{i+1} \right|^{2}\right)}} \right]^{\bigast}.
  \end{equation}
\end{subequations}
In the second quantization language $\xi_{i}$ ($\bar{\xi}_{i}$) creates (annihilates) a spinless fermion at $i$-th site. Similarly, $z_{i}$ ($\bar{z}_{i}$) creates (annihilates) a spinfull boson at $i$-th site. In Eq. (\ref{eq:effective-ham-cov}) the $\xi$ and $z$ degrees of freedom emerge as low-energy degrees of freedom.

One can absorb the electron hopping and RSOC terms in Eq. (\ref{eq:effective-ham-cov}) into a single effective term. In this case the Hamiltonian, Eq. (\ref{eq:effective-ham-cov}), transforms to:
\begin{equation}
  \label{eq:effectiv-ham-trans}
  \begin{aligned}
    H_{\text{eff}}(z, \xi) =& - \sum\limits_{i} \bar{\xi}_{i} \xi_{i+1} \left(t \: a_{i,i+1}^{\ast} + \imath \alpha \: \alpha_{i,i+1} \right)  \\
    &\quad - \Delta\sum\limits_{i}\bar{\xi}_{i}\bar{\xi}_{i+1} \Delta_{i,i+1} + \text{H.c.} + \mu \sum\limits_{i} \bar{\xi}_{i} \xi_{i}.    
  \end{aligned}
\end{equation}
In effect, electron hopping acquires a phase which is a complex function of the spin field. One can define the modified hopping as:
\begin{equation}
  \label{eq:modified-hopping-general}
  \tilde{t}_{i} = \sqrt{ \left(t \: a_{i,i+1}^{\ast} \right)^{2} +  \left(  \alpha \: \alpha_{i,i+1}\right)^{2} } \: \exp \left[ \imath \: \arctan \left( \frac{\alpha \: \alpha_{i,i+1}}{t \: a_{i,i+1}^{\ast} } \right) \right].
\end{equation}

\section{Hamiltonian for spiral spin fields}
\label{sec:emerg-topol-effects}
While treating the real systems, it is instructive physically to treat the dynamic spin fields $z_{i}(t)$ in terms of the spin covariant symbols \cite{ferraz_2011_EffectiveAction_NuclearPhysicsB}:
\begin{equation}
  \label{eq:spin-cov-symb}
  \begin{aligned}
    S_{i}^{+} = \frac{z_{i}}{1+\left| z_{i} \right|^{2}},\quad
    S_{i}^{-} = \frac{\bar{z}_{i}}{1+\left| z_{i} \right|^{2}},\quad
    S_{i}^{z} = \frac{1}{2} \left( \frac{1- \left| z_{i} \right|^{2}}{1+\left| z_{i} \right|^{2}} \right).
  \end{aligned}
\end{equation}
The corresponding expressions for $S_{i}^{x}=S_{i}^{+}+S_{i}^{-}$ and $S_{i}^{y}=-\imath \left(S_{i}^{x} -S_{i}^{y} \right)$ can be easily found. In strongly correlated electronic systems the dynamics of the effective spin field is much slower than that of the fermionic fields \cite{herbrych_2021_InteractioninducedTopological_NatCommun}. Because of this we can apply the mean-field approximation to the spin system (from the point of view of the fermions). In other words, the picture where spinless fermions travel in the static background of the spin field is fully applicable. One of the main difference between our approach and the previous works related to the 1D topological superconductors \cite{sau_2010_RobustnessMajorana_PhysRevB,oreg_2010_HelicalLiquids_PhysRevLett,nadj-perge_2013_ProposalRealizing_PhysRevB,choy_2011_MajoranaFermions_PhysRevB} is that, we first derive our effective Hamiltonian taking full account of the strong correlation effects and only then we apply mean-field approximation. As opposed to that in previous works the mean-field approximation is applied to the Hamiltonian right from the beginning. In fact using density matrix renormalization group it was explicitly shown that usual mean field models fails when strong correlation is present \cite{aksenov_2020_StrongCoulomb_PhysRevB}.

We consider a spin field of the form:
\begin{equation}
  \label{eq:spin-config-simp}
  \vec{S}_{i} = \left( S_{i}^{x}, S_{i}^{y}, S_{i}^{z} \right) = \frac{1}{2}\left(\cos \theta_{i}, \sin \theta_{i}, 0  \right).
\end{equation}
Physically it represents the rotating spin vector on the $xy$ plane; $z$ component of the spin is zero. $\theta_{i}$ is the angle on the \emph{xy} plane. It is related to the spatial coordinate through relation: $\theta_{i}=\vec{q} \cdot \vec{r}_{i}$; with $\vec{q}$ being the spin wave vector and $\vec{r}_{i}$ is the position vector. Substituting Eq. (\ref{eq:spin-config-simp}) in Eq. (\ref{eq:spin-cov-symb}) it is easy to see that $z_{i}=e^{\imath \theta_{i}}$. Further, using this $z_{i}$ in Eq. (\ref{eq:def-aij}), (\ref{eq:def-alpha-ij}), and (\ref{eq:def-delta-ij}) one will find:
\begin{equation}
  \label{eq:spin-config-simp-aij-alpij-delij}
  \begin{aligned}
    &a_{i,i+1}^{\ast} &&= \quad \mathrm{e}^{-\imath \left(\theta_{i+1}-\theta_{i}\right)/2} \quad \cos \frac{\left(\theta_{i+1}-\theta_{i} \right)}{2},\\
    &\alpha_{i,i+1} &&=  \: \: \: \imath \mathrm{e}^{-\imath \left(\theta_{i+1}-\theta_{i}\right)/2} \quad \sin \frac{\left(\theta_{i+1}+\theta_{i} \right)}{2},\\
    &\Delta_{i,i+1} &&=  -\imath \mathrm{e}^{\imath \left(\theta_{i+1}-\theta_{i}\right)/2} \quad \: \: \sin \frac{\left(\theta_{i+1}-\theta_{i} \right)}{2} \: \mathrm{e}^{\imath \theta_{i}}.
  \end{aligned}
\end{equation}
Substituting Eq. (\ref{eq:spin-config-simp-aij-alpij-delij}) in Eq. (\ref{eq:effectiv-ham-trans}), and soon after that making the gauge transformations $\xi_{i} \to \xi_{i} \mathrm{e}^{\imath \left( \theta_{i}/2\: - \: \pi/4 \right)}$ the effective Hamiltonian can be written as:
\begin{equation}
  \label{eq:effective-ham-simp-spin-config}
  \begin{aligned}
    H_{\text{eff}} = & - t \sum\limits_{i} \bar{\xi}_{i} \xi_{i+1} \cos \frac{\left(\theta_{i+1}-\theta_{i} \right)}{2}\\
                             &\quad + \alpha \sum\limits_{i} \bar{\xi}_{i} \xi_{i+1} \sin \frac{\left(\theta_{i+1}+\theta_{i} \right)}{2}\\
                             &\qquad - \Delta\sum\limits_{i}\bar{\xi}_{i}\bar{\xi}_{i+1} \sin \frac{\left(\theta_{i+1}-\theta_{i} \right)}{2} + \text{H.c.} \\
                             & \qquad \quad+ \mu \sum\limits_{i} \bar{\xi}_{i} \xi_{i}.
  \end{aligned}
\end{equation}

\begin{figure}
  \centering
   \includegraphics[width=.45\textwidth]{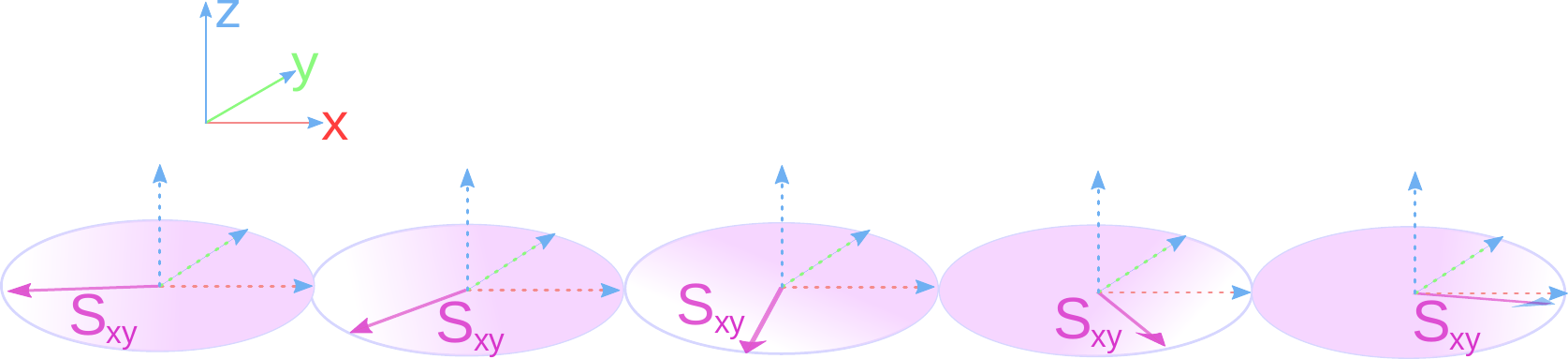}
   \caption{Spiral spin structure found from Eq. (\ref{eq:spin-config-simp}) by taking $\theta_{i+1}-\theta_{i}=\theta$ as constant.}
  \label{fig:spiral-spin-struct}
\end{figure}
In a limiting case one can assume that the rotation of the spin through neighboring sites $\theta_{i+1}-\theta_{i} \equiv \theta$ is constant. This spin structure is known as the spiral spin structure. A schematic of it is shown in Fig. \ref{fig:spiral-spin-struct}. The Hamiltonian for the spiral spin structure then becomes:
\begin{equation}
  \label{eq:effective-ham-simp-spin-config-constant-theta}
  \begin{aligned}
    H_{\text{eff}} = & - t \sum\limits_{i} \bar{\xi}_{i} \xi_{i+1} \cos \frac{\theta}{2} + \alpha \sum\limits_{i} \bar{\xi}_{i} \xi_{i+1} \sin \left( \theta_{i} +\frac{ \theta}{2} \right)\\
                   &\quad - \Delta\sum\limits_{i}\bar{\xi}_{i}\bar{\xi}_{i+1} \sin \frac{\theta}{2} + \text{H.c.} + \mu \sum\limits_{i} \bar{\xi}_{i} \xi_{i}.
  \end{aligned}
\end{equation}
The Rashba term is now real and oscillating.

Before further discussion, a explanation of why we considered the spiral spin texture is necessary. The spiral spin texture is not new and it has been observed in other systems, especially, in heterostructures where magnetic ad-atoms are placed over superconducting substrates \cite{nadj-perge_2013_ProposalRealizing_PhysRevB,nadj-perge_2014_ObservationMajorana_Science}. In these systems strong RSOC \cite{li_2014_TopologicalSuperconductivity_PhysRevB}, together with RKKY interaction \cite{klinovaja_2013_TopologicalSuperconductivity_PhysRevLett} are responsible for generating spiral spin texture. In fact, it can be shown that the tight binding Hamiltonian of a 1D wire with spin-orbit coupling is equivalent to the tight binding Hamiltonian with a rotating magnetic field \cite{braunecker_2010_SpinselectivePeierls_PhysRevB}. This result is valid only when the \emph{e-e} interaction is small ($U \ll t$). When the \emph{e-e} interaction is strong the Hamiltonian cannot be transformed in this way. This is due to  the fractionalization of the electronic degrees of freedoms into separate spinless fermionic field (charge) and spinful bosonic field (spin).

The mechanisms for emergence of the spiral spin texture in case of weak interactions are not applicable to the case of strong correlations. Therefore we suggest the following mechanisms and approximations for the emergence of the spiral spin texture. For the strong \emph{e-e} interaction nearly localized effective spins emerge at individual sites. The effective spinful bosonic field is generated by Hubbard operators. A crucial observation being that the operators $X^{\sigma 0}$  transform in the spinor (fundamental) representation of the $SU(2)$ group \cite{scheunert_1977_GradedLie_JMathPhys}. Because of this, the local spin operators that describe the effective spin degrees of freedom of strongly correlated electrons can be written as:
\begin{equation}
  \label{eq:q-representation}
   \vec Q_i:= X^{\dagger}_i\vec \sigma X_i,
\end{equation}
where $X_i:= \left(X^{0 \uparrow }_i, X^{0 \downarrow }_i\right)^T$, It can be checked that a set of the spin operators $\vec{Q}$ spans the $su(2)$ algebra. The explicit coherent-state symbols of those operators can be found in Ref. \cite{ferraz_2022_FractionalizationStrongly_PhysRevB}. Physically we have the effective system where at each lattice site sits a classical spin with $S=1/2$, and in this spin background itinerant spinless fermions travel through the lattice. The nearly localized spins create a dynamical spin field for which the internal energy of the system is lowest.

Since the background spin field is generated dynamically, it is natural to assume that, the system chooses the spin configuration for which the internal energy is lowest. Keeping this in mind, for analysis of the properties of the Hamiltonian, one can start with any non-colinear spin structure, however, for a general complex spin textures --- e.g. for skyrmions, hopfions etc --- the resulting Hamiltonian, Eq. (\ref{eq:effectiv-ham-trans}), becomes too complex to be solved analytically. Hence, some assumptions are in order. First, we assume that the \emph{z} component of the spin is zero; physically, it means that the spin rotates only on the $xy$ plane. This was motivated by the fact that the spin projection along \emph{z}-axis does not play any role in determining the topological properties of the system for not too large spins $S$ \cite{kesharpu_2023_FlippingChiral_}; this means that only the spin rotations on the $xy$ plane are important. Secondly, we assume that the rotation of the spin projection on the $xy$ plane between two neighbouring sites is constant. It is motivated by two facts: (i) a constant rotation angle --- which corresponds to the spiral spin texture --- has been observed in numerous experiments \cite{lutchyn_2018_MajoranaZero_NatRevMater,nadj-perge_2013_ProposalRealizing_PhysRevB,nadj-perge_2014_ObservationMajorana_Science}, (ii) in this case the Hamiltonian takes a very simple form, which turns out to be solvable analytically.

\section{Absence of RSOC ($\alpha=0$)}
\label{sec:absence-rsoc-alpha=0}
\begin{figure}
  \centering
  \subfloat[]{\label{fig:kitaev-phase-diag-a}\includegraphics[width=.24\textwidth]{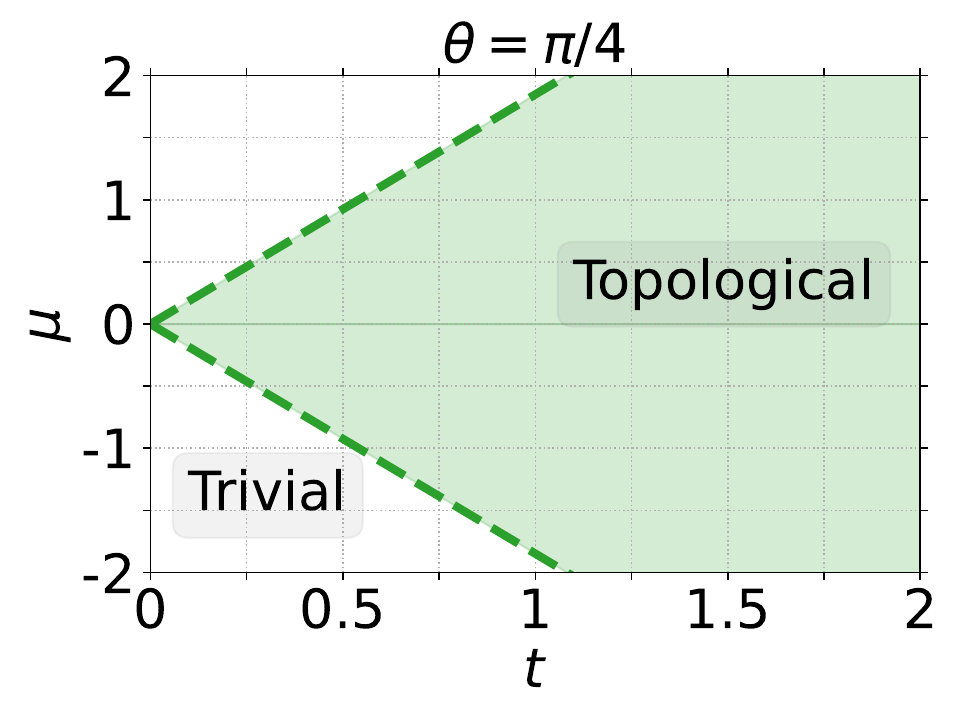}}
  \subfloat[]{\label{fig:kitaev-phase-diag-b}\includegraphics[width=.24\textwidth]{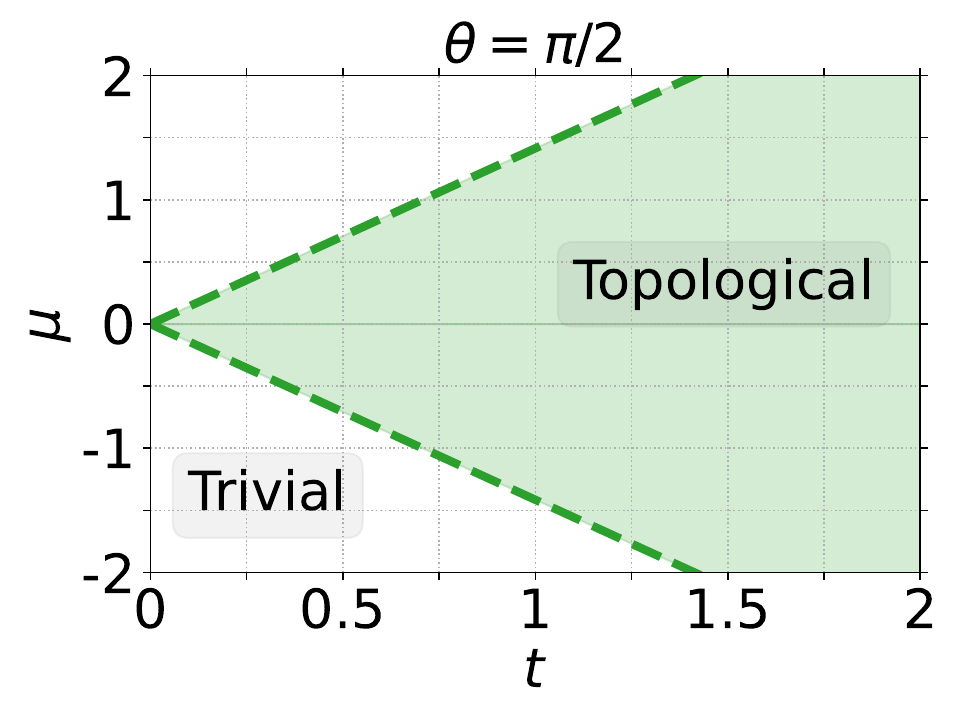}}\\
  \subfloat[]{\label{fig:kitaev-phase-diag-c}\includegraphics[width=.24\textwidth]{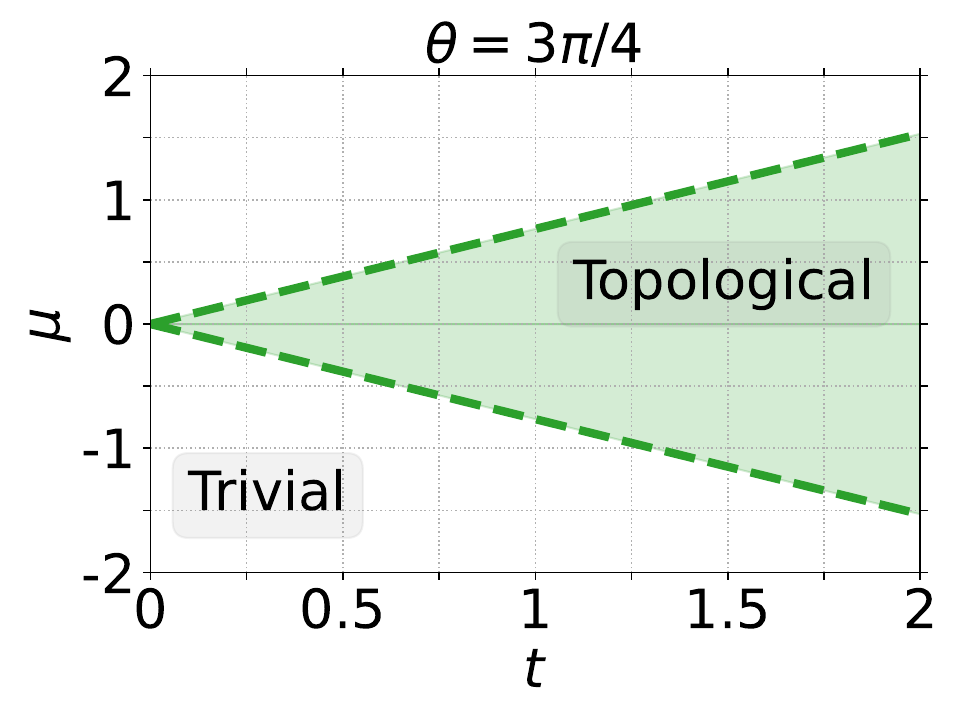}}
  \subfloat[]{\label{fig:kitaev-phase-diag-d}\includegraphics[width=.24\textwidth]{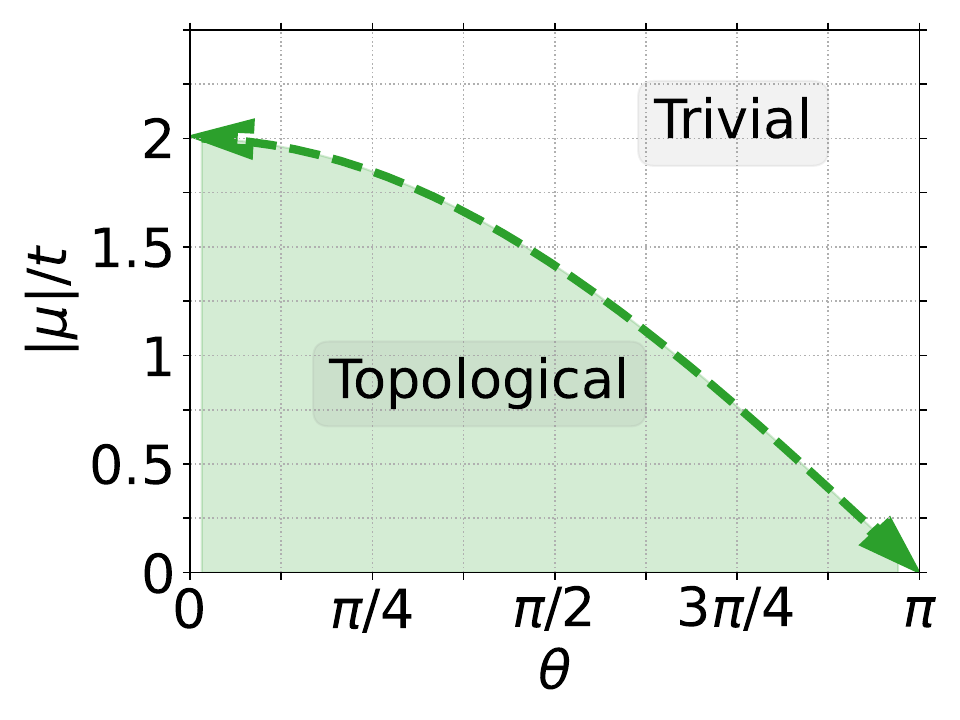}}\\
  \caption{(a,b,c) Dependence of trivial and topological phase on the angle of the rotation of the spin through neighboring sites ($\theta$). The topological phase appears when $\left| \mu \right|< 2 t \cos \left( \theta/2 \right)$ is satisfied. The range of parameters $\mu$ and $t$, for which topological phase occurs, decreases with increase in $\theta$. (d) The condition $\left| \mu/t \right|<2 \cos \left( \theta/2 \right)$ determining the topological phase is plotted. The area for the topological phase in parameter space is largest when spins on neighbouring atoms are almost aligned along same direction ($\theta > 0$). For ferromagnetic ordering ($\theta=0$) the topological phase vanishes as the superconducting gap $\left[\Delta \sin \left( \theta/2 \right)\right]$ also vanishes. For anti-ferromagnetic ordering ($\theta=\pi$) the topological phase vanishes, because electron hopping becomes zero $\left[t \cos \left( \theta/2 \right) \right]$. The arrow marks on the boundary of the topological and trivial phase near $\theta=0,\pi$ represent the fact that, topological phase survives as $\theta \to 0$ and $\theta \to \pi$, however at $\theta=0,\pi$ topological phase is absent.}
  \label{fig:kitaev-phase-diag}
\end{figure}
First, we analyze the Eq. (\ref{eq:effective-ham-simp-spin-config-constant-theta}) when the RSOC is absent, i.e. we take $\alpha=0$. The effective Hamiltonian in this case becomes:
\begin{equation}
  \label{eq:effective-ham-kitaev}
  \begin{aligned}
    H^{\text{Kitaev}}_{\text{eff}} = & - t \sum\limits_{i} \bar{\xi}_{i} \xi_{i+1} \cos \frac{\theta}{2} - \Delta\sum\limits_{i}\bar{\xi}_{i}\bar{\xi}_{i+1} \sin \frac{\theta}{2}\\
                                 &\qquad+ \text{H.c.} + \mu \sum\limits_{i} \bar{\xi}_{i} \xi_{i}.
  \end{aligned}
\end{equation}
This Hamiltonian does not break time reversal symmetry, as $ H^{\text{K}}_{\text{eff}} = \left(H^{\text{K}}_{\text{eff}} \right)^{\ast}$. In this sense it is analogous to the Kitaev's toy model of topological superconductivity \cite{kitaev_2001_UnpairedMajorana_Phys-Usp}\footnote{The Kitaev model is represented as:
  \begin{equation*}
    \label{eq:kitaev-model}
    H^{\text{K}} = \sum\limits_{i} -t f_{i} f^{\dagger}_{i+1} + \Delta f^{\dagger}_{i}  f^{\dagger}_{i+1} + \text{H.c.} + \mu f_{i}f_{i}^{\dagger}.
  \end{equation*}
  It is also symmetric under time reversal transformation.
}. In fact, analogous to the Kitaev's toy model, we also find that, both the trivial and the topological phases depend on the specific values of $t$, $\Delta$, $\mu$, and $\theta$. When there is no superconductivity, the system is gapped for $\left| \mu \right| > 2 t \cos (\theta/2)$; but it is gapless when $\left| \mu \right| < 2 t \cos (\theta/2)$. However when superconductivity is itself present, due to the particle hole symmetry, a topological phase appears for $\left| \mu \right| < 2 t \cos (\theta/2)$. The phase diagram of the $H_{\text{eff}}^{\text{Kitaev}}$, and its dependence on the $\theta$ --- physically, $\theta$ represents the rotation of the spin projection through neighboring sites --- is shown in Fig. \ref{fig:kitaev-phase-diag}. The main difference between the phase diagrams shown in Fig. \ref{fig:kitaev-phase-diag}, and the original phase diagram of the Kitaev chain \cite{kitaev_2001_UnpairedMajorana_Phys-Usp}, is the emergence of the new parameter $\theta$ in the effective Hamiltonian. It can be observed that with the increase in $\theta$ the area of the topological phase in the $\mu$-$t$ phase plane decreases. This is due to the decrease in the electron hopping parameter $t \cos \left( \theta/2 \right)$. The boundary between the topological and the trivial phase in the  $\left|\mu\right|/t$-$\theta$ phase space is given by the condition $\left| \mu \right|/t <2 \cos (\theta/2)$. The corresponding phase diagram is shown in Fig. \ref{fig:kitaev-phase-diag-d} . In this figure we observe that, when $\theta=0$, i.e. for the ferromagnetic state, the system behaves either as a simple metal or as an insulator depending on the relative value of $\mu$ \footnote{If $-2t \cos \left( \theta/2 \right) \cos k <\mu<0$ then, due to unfilled band the system behaves as metal. However, if  $\mu \approx 0$ and the Fermi energy lies inside the band gap, and the system behaves as insulator.}. The reason, why the system looses its topological property at $\theta=0$, is due to the disappearance of the superconducting gap. In effect, the system looses the particle-hole symmetry, which in the first place gives rise to the topological phase. This is expected physically since the underlying extended $s$-wave superconducting state should not survive in a ferromagnetic background. On the other end when $\theta=\pi$, i.e. for an antiferromagnetic state, the system behaves as an usual Mott insulator. The hopping of the electrons becomes identically zero and this represents a state in which the fermionic particle are frozen in their lattice sites. From the band theory point of view, the superconducting gap never closes even if $\mu$ is modified, when $\theta=\pi$. This means that the MZM can not be created; since in such a state Majorana fermions only emerge during the closing and the reopening of the band gap.

\begin{figure}
  \centering
   \includegraphics[width=.48\textwidth]{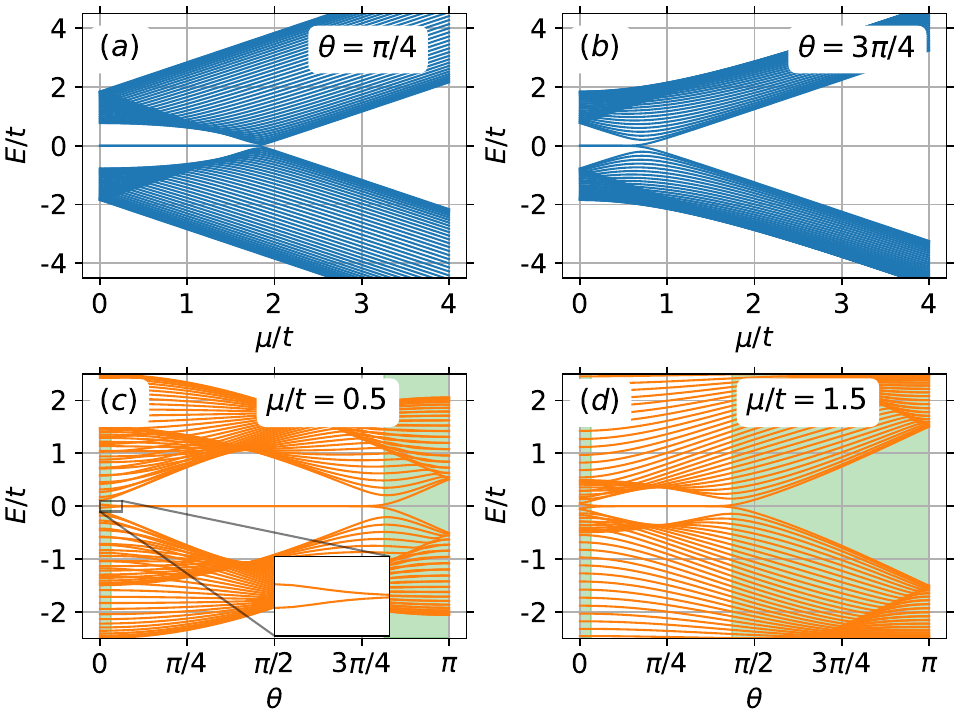}
  \caption{Dependence of real space energy modes on $\mu$ and $\theta$ for a chain of 35 atoms. (a, b) Energy modes are calculated by numerically diagonalizing the Hamiltonian Eq. (\ref{eq:effective-ham-kitaev}) for $\theta=\pi/4$ and $\theta=3 \pi/4$. We fixed the superconducting gap $\Delta=0.5t$. The zero-mode energy is present upto values of $\mu/t=2 \cos \left( \theta/2 \right)$. (c, d) Dependence of energy modes on $\theta$ for $\mu=0.5t$ and $\mu=1.5t$. The inset in Fig. \ref{fig:energy-spectrum}c shows the absence of zero mode energy at $\theta \to 0$.}
  \label{fig:energy-spectrum}
\end{figure}
To support the previous results we numerically diagonalize the real space Hamiltonian Eq. (\ref{eq:effective-ham-kitaev}) for a chain of 35 atoms. First the energy modes are calculated for different $\mu$ keeping $\theta=\pi/4$ (Fig. \ref{fig:energy-spectrum}a) and $\theta=3\pi/4$ (Fig. \ref{fig:energy-spectrum}b) constant. One can clearly see the zero mode energy and their splitting as $\mu$ increases. In Fig. \ref{fig:energy-spectrum}c and \ref{fig:energy-spectrum}d we analyze the dependence of the energy modes on $\theta$ while chemical potential $\mu $ is kept constant. In these figures one can see that when $\theta \to 0$ and $\theta = \pi$ zero mode energy is absent as was predicted in Fig. \ref{fig:kitaev-phase-diag}d. If we compare Fig. \ref{fig:energy-spectrum}c,d with Fig. 2a of Ref. \cite{nadj-perge_2013_ProposalRealizing_PhysRevB}, then we find an analogous dependence on the magnetic field; in our case $\theta$ plays the same role as the magnetic field in Ref. \cite{nadj-perge_2013_ProposalRealizing_PhysRevB}.

\begin{figure}
  \centering
\subfloat[]{\label{fig:energy-disp-const-delta-a}\includegraphics[width=.24\textwidth]{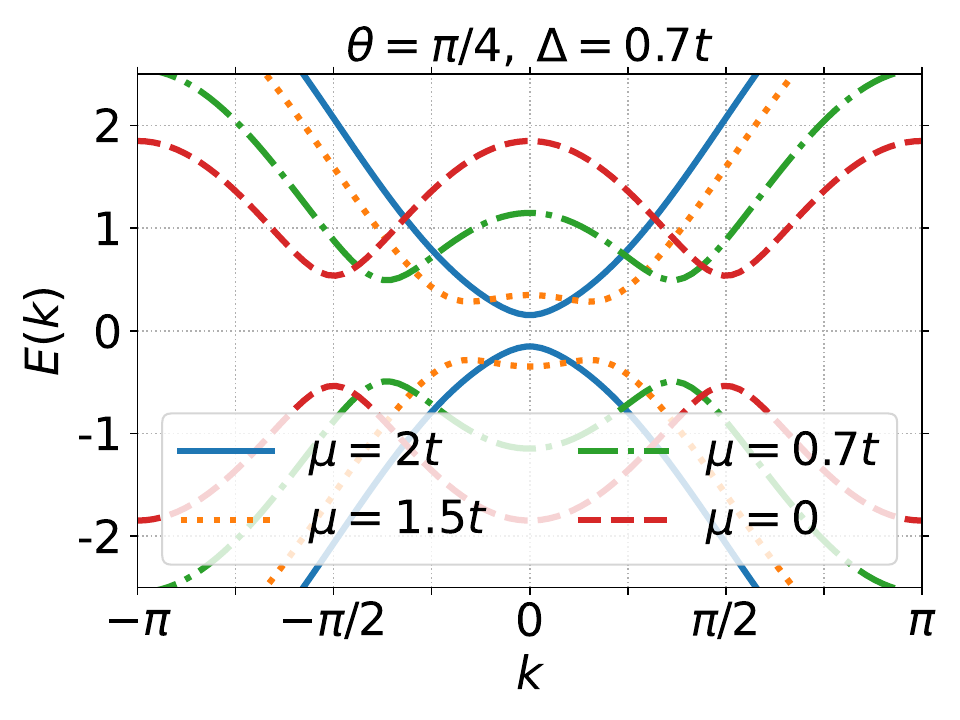}}
  \subfloat[]{\label{fig:energy-disp-const-delta-b}\includegraphics[width=.24\textwidth]{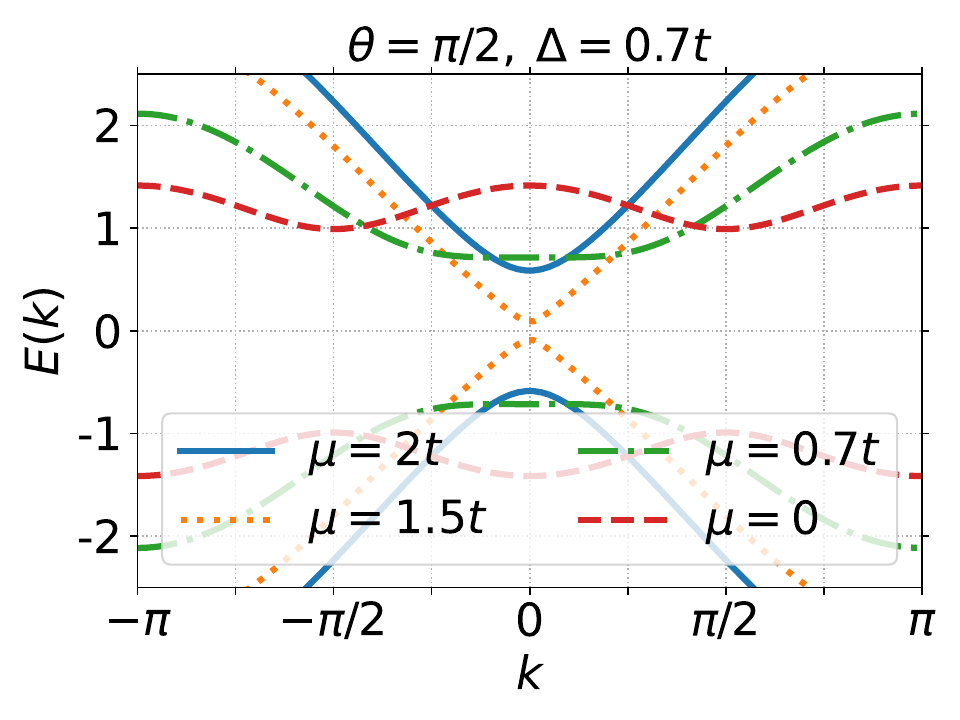}}\\
  \subfloat[]{\label{fig:energy-disp-const-delta-c}\includegraphics[width=.24\textwidth]{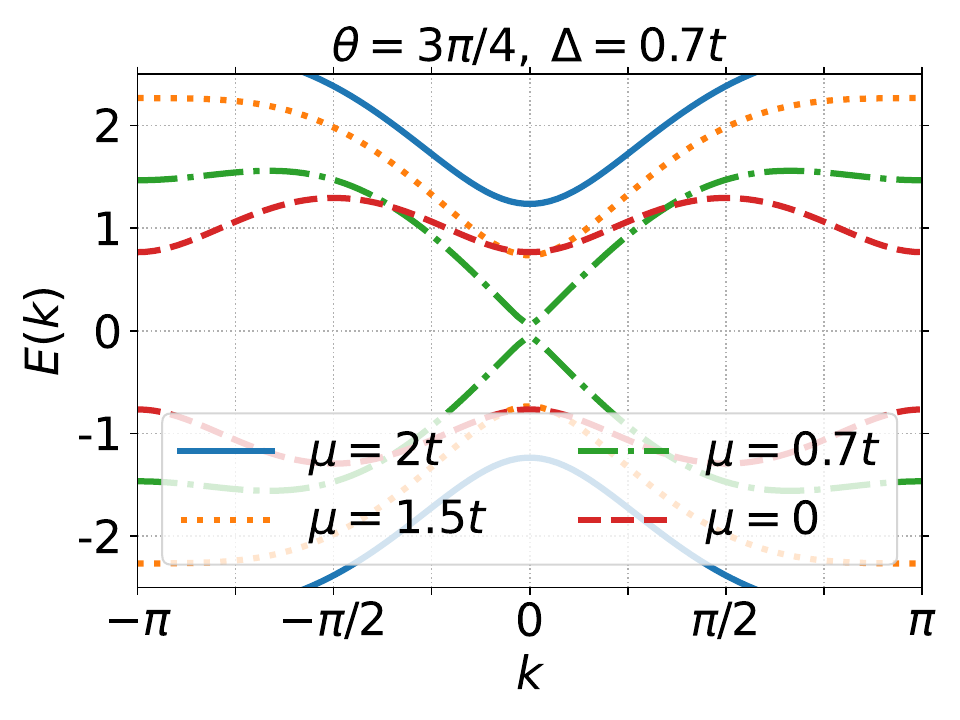}}
  \subfloat[]{\label{fig:energy-disp-const-delta-d}\includegraphics[width=.24\textwidth]{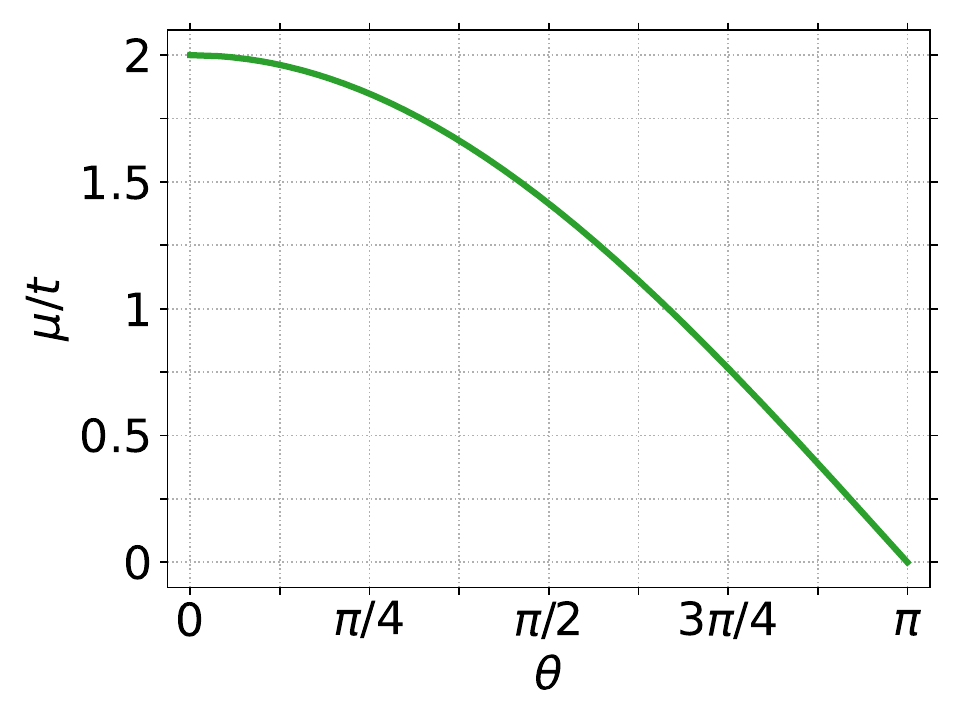}}\\
  \caption{(a,b,c) Evolution of the energy dispersion, Eq. (\ref{eq:energ-disp-eff-Kit-Ham}), as the superconducting gap is kept constant ($\Delta=0.7t$), and chemical potential is decreased ($\mu$ changes from $2t$ to $0$) for $\theta=\pi/4$, $\theta=\pi/2$, and $\theta=3\pi/4$. (d) The values of $|\mu|/t$ at which the bulk gap closes at $k=0$ for a given value of $\theta$. The values are found by using the condition $\theta = 2 \arccos \left( |\mu|/2t \right)$.}
  \label{fig:energy-disp-const-delta}
\end{figure}
\begin{figure}
  \centering
\subfloat[]{\label{fig:energy-disp-const-mu-a}\includegraphics[width=.24\textwidth]{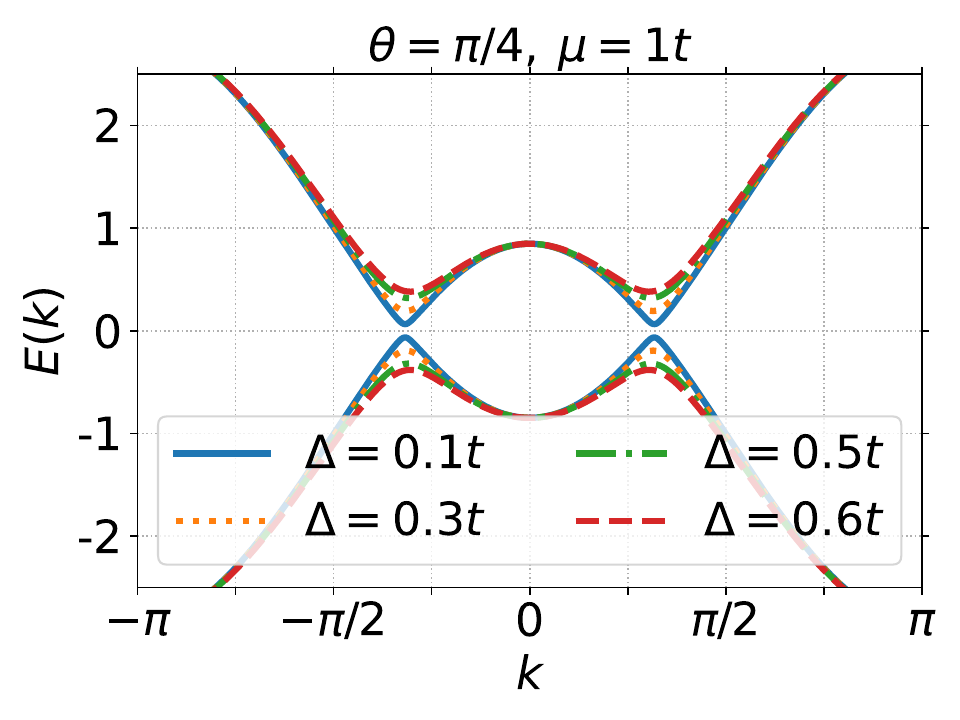}}
  \subfloat[]{\label{fig:energy-disp-const-mu-b}\includegraphics[width=.24\textwidth]{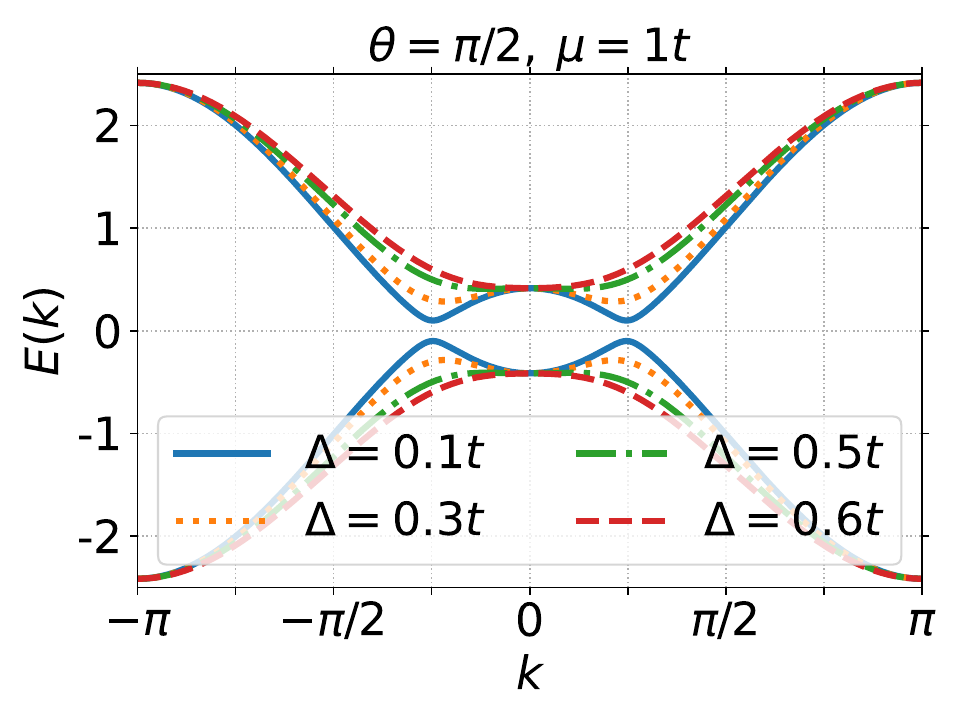}}\\
  \subfloat[]{\label{fig:energy-disp-const-mu-c}\includegraphics[width=.24\textwidth]{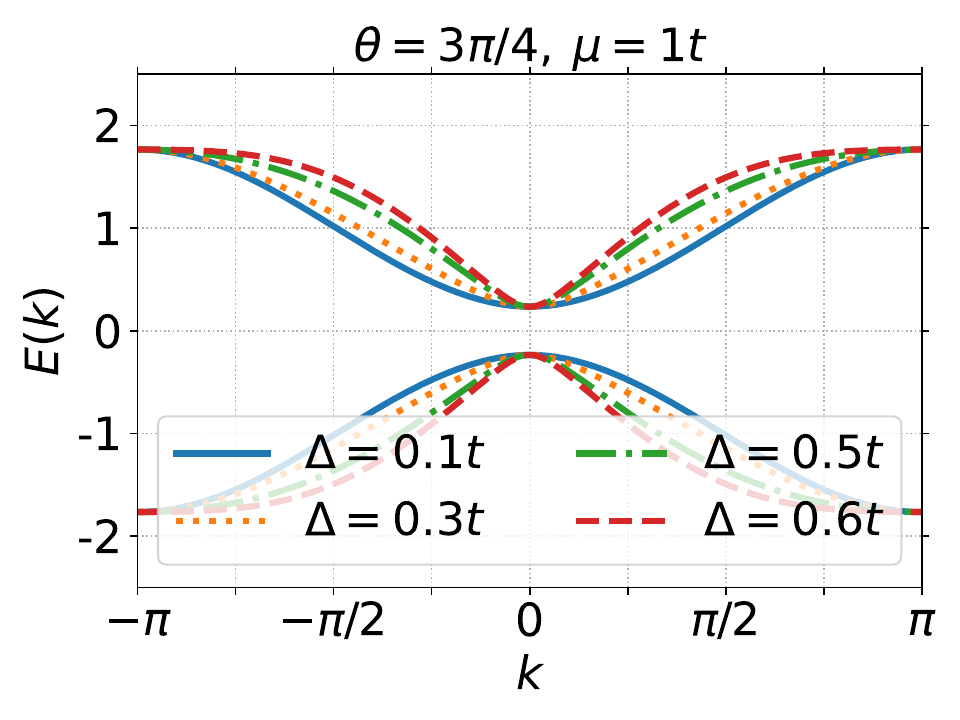}}
  \subfloat[]{\label{fig:energy-disp-const-mu-d}\includegraphics[width=.24\textwidth]{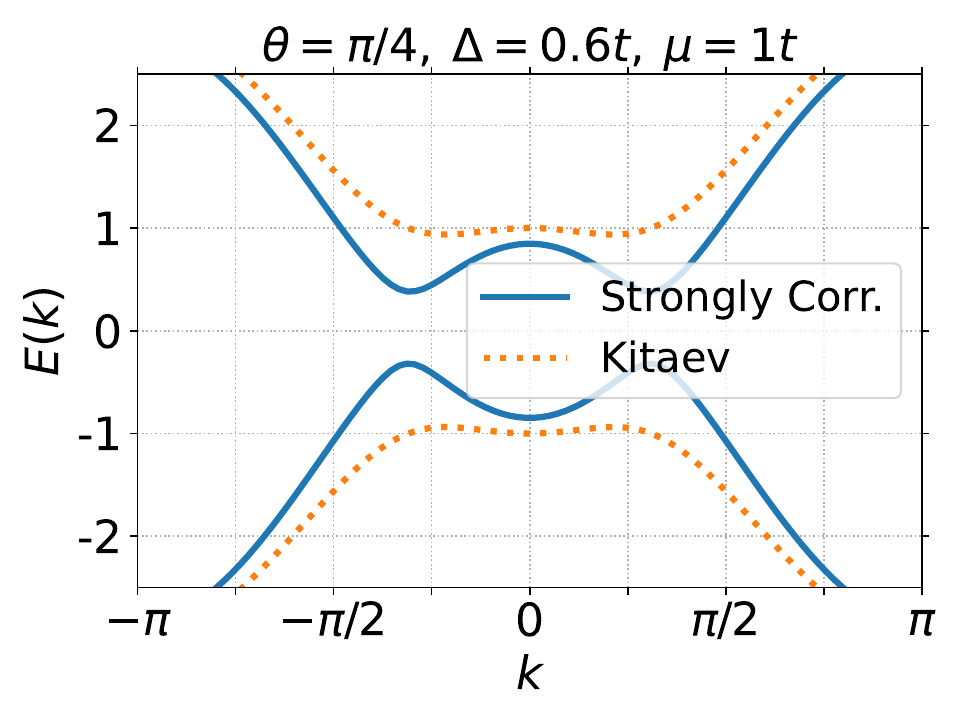}}\\
  \caption{(a,b,c) Evolution of the energy dispersion, Eq. (\ref{eq:energ-disp-eff-Kit-Ham}), as the chemical potential is kept constant ($\mu=1t$), and the superconducting gap is increased ($\Delta=0.1t, 0.3t, 0.5t, 0.6t$) for $\theta=\pi/4$, $\theta=\pi/2$, and $\theta=3\pi/4$. (d) Comparison of the energy dispersion for the strongly correlated Hamiltonian, Eq. (\ref{eq:effective-ham-kitaev}), and the Kitaev's toy model \cite{kitaev_2001_UnpairedMajorana_Phys-Usp} at $\mu=1t$, $\Delta=0.6t$, and $\theta=\pi/4$.}
  \label{fig:energy-disp-const-mu}
\end{figure}

Next, we discuss the effect of $\theta$ on the energy spectrum. Using Nambu spinors in momentum space the Hamitonian Eq. (\ref{eq:effective-ham-kitaev}) can be written in the usual Bogoliubov-de Gennes form:
\begin{equation}
  \label{eq:effectiv-Ham-Kit-BdG}
  \begin{aligned}
    &H_{\text{eff}}^{\text{Kitaev}}(k) = \frac{1}{2} \sum\limits_{k \in \text{BZ}} \Psi_{k}^{\dagger} \: \mathcal{H}_{k} \: \Psi_{k},\\
    &\text{where,}\\
    & \Psi_{k}^{\dagger} \equiv \left[ \xi_{k}, \bar{\xi}_{-k} \right],\\
    &  \mathcal{H}_{k} \equiv
      \begin{bmatrix}
        2t \cos \frac{\theta}{2} \cos ka - \mu  & 2\Delta  \sin \frac{\theta}{2} \sin ka \\
        2\Delta \sin \frac{\theta}{2} \sin ka   & 2t \cos \frac{\theta}{2} \cos ka - \mu \\
      \end{bmatrix}.
  \end{aligned}
\end{equation}
Here, $k$ is the crystal momentum, and $a$ is the distance between two near neighbour lattice sites.
The single mode energy dispersion can now be easily calculated:
\begin{equation}
  \label{eq:energ-disp-eff-Kit-Ham}
  \begin{aligned}
    E(k) = \pm \sqrt{\left( 2t \cos \frac{\theta}{2} \cos ka - \mu \right)^{2} + \left( 2\Delta  \sin \frac{\theta}{2} \sin ka \right)^{2}}.
  \end{aligned}
\end{equation}
If one compares the energy dispersion of Eq. (\ref{eq:energ-disp-eff-Kit-Ham}), with the energy dispersion of the Kitaev's toy model \cite[see Eq. (13) of Ref. ][]{kitaev_2001_UnpairedMajorana_Phys-Usp} again, there is only single difference; in Eq. (\ref{eq:energ-disp-eff-Kit-Ham}) the extra parameter $\theta$ appears explicitly in our energy dispersion. The dependence of energy dispersion for different combination of the $\Delta$, $\mu$ and $\theta$ is shown in Fig. \ref{fig:energy-disp-const-delta} and \ref{fig:energy-disp-const-mu}.

In Figs. \ref{fig:energy-disp-const-delta-a}, \ref{fig:energy-disp-const-delta-b}, and \ref{fig:energy-disp-const-delta-c} the evolution of the energy dispersion are shown for $\theta=\pi/4$, $\pi/2$, and $3\pi/4$, while the superconducting gap is kept constant $(\Delta=0.7t)$, and the chemical potential is varied as a function of $t$ $(\mu=2t$, $1.5t$, $0.7t$, $0)$. As is well known one of the properties of the topological superconductors is its sensitivity to the closing and to the reopening of the energy gap at $k=0$, $\pm \pi$ when we tune $\mu$. In the Kitaev's toy model this takes place precisely at $\mu=\pm 2t$. However, in our case the gap closes for $\mu=\pm 2t \cos \left(\theta/2\right)$ --- it evolves with $\theta$. The values of $\mu/t$ where the gap closes, for a given $\theta$ is plotted in Fig. \ref{fig:energy-disp-const-delta-d}. Similarly, in Figs. \ref{fig:energy-disp-const-mu-a}, \ref{fig:energy-disp-const-mu-b}, and \ref{fig:energy-disp-const-mu-c} the energy dispersion are shown for $\theta=\pi/4$, $\pi/2$, and $3\pi/4$, while the chemical potential is kept constant $(\mu=1t)$, and the superconducting gap is varied $(\Delta=0.1t$, $0.3t$, $0.5t$, $0.6t)$. As expected for $\theta = \pi/4$ the spectrum is defined by the electron hopping term $\left[2t \cos \left( \theta/2 \right) \cos k \right]$; the superconducting term has little effect as $\left[2t \sin \left( \theta/2 \right) \sin k \right]$ is very small. However, for $\theta = 3\pi/4$ the spectrum is defined by the superconducting term.

\begin{figure}
  \centering
   \includegraphics[width=.48\textwidth]{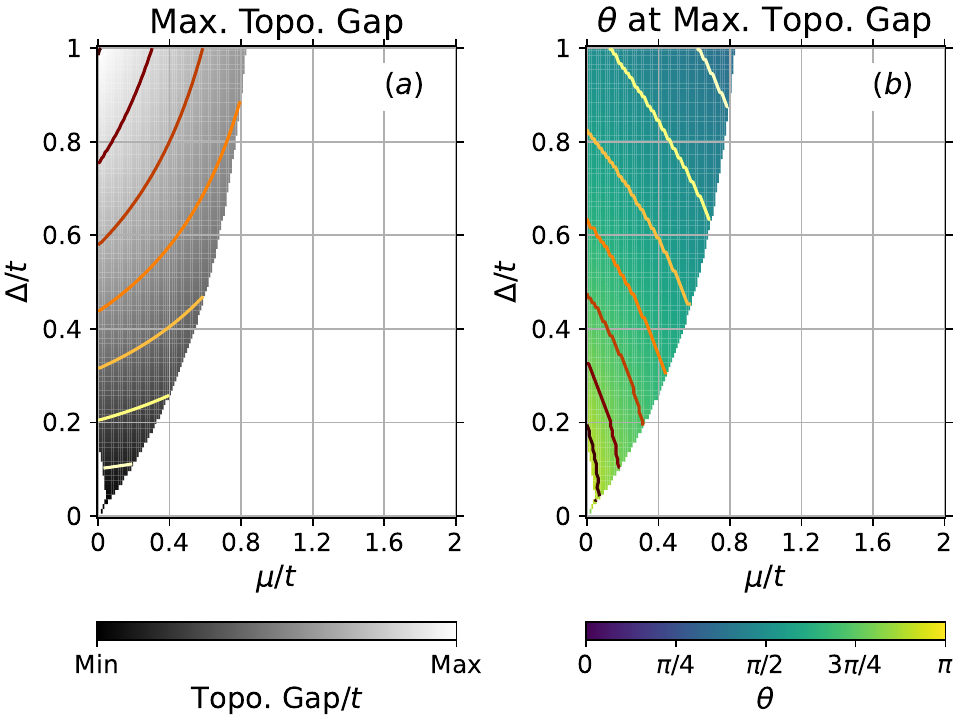}
  \caption{Dependence of the topological gap on $\theta$, $\mu$, $\Delta$ calculated numerically using dispersion Eq. (\ref{eq:energ-disp-eff-Kit-Ham}). (a) Values of the maximum topological gap energy for different $\mu$ and $\Delta$. (b) Corresponding values of $\theta$ for which maximum topological gap occurs. The lines show the contour line of equal values of the topological gap (a) and $\theta$ (b).}
  \label{fig:mu-delta-large-gap}
\end{figure}
For practical application one is always interested in increasing the topological gap in the bandstructure. As the dispersion is a function of $\theta$, $\mu$, and $\Delta$, the topological gap will also depend on these parameters. In Fig. \ref{fig:mu-delta-large-gap} we showed the highest topological gap and corresponding $\theta$ for a given value of $\mu$ and $\Delta$. In Fig. \ref{fig:mu-delta-large-gap}b we show the value of $\theta$ for which the maximum topological gap results; the corresponding topological gap energy is shown in Fig. \ref{fig:mu-delta-large-gap}a. In both these figures we didn't show the value of $\mu$ and $\Delta$ for which $\theta=0, \pi$, as at this values of $\theta$  there is no topological phase. Regarding the optimal parameter for achieving highest topological gap is when $\mu \lessapprox 0.8t$; the dependence on $\Delta$ is expected, as increase in $\Delta$ will give higher topological gap.

The angle $\theta=\vec{q} \cdot (\vec{r}_i-\vec{r}_{i+1})=\vec{q} \cdot \vec{a}$ where $\vec{a}$ is a lattice spacing and $\vec{q}$ is a spin spiral modulation vector appears as a control parameter. It can in principle be tuned by changing $a$. However such a change affects the other parameters in the Hamiltonian as well. One could alternatively try to change the modulation vector $\vec{q}$  by applying an external modulated field as discussed in Sec. \ref{sec:prop-exper-heter}. It is still however not quite clear how the tunability of $\theta$ could  be realized in an efficient way.

A comment is in order regarding the related work of Ref. \cite{ferraz_2023_ConnectionKitaev_AnnalsOfPhysics}. If one compares our Hamiltonian, Eq. (\ref{eq:effectiv-Ham-Kit-BdG}) with Eq. (20) of the Ref. \cite{ferraz_2023_ConnectionKitaev_AnnalsOfPhysics}, they are identical. However, there are a few fundamental differences in what the two models truly represent. In Ref. \cite{ferraz_2023_ConnectionKitaev_AnnalsOfPhysics} the heterostructures are produced by the magnetic ad-atoms on a superconductor, while in our work we consider a strongly correlated 1D nanowire on the superconductor. In the magnetic ad-atom superconductor heterostructures the whole physics is directly related to the YSR states, while in our system the physics depends on the proximity effect induced by the superconductivity and by the chemical potential of the nanowires. In this sense they are completely different systems. A priori it is not obvious they can be represented by one and the same low-energy theory. This only happens in the limit where on-site Coulomb repulsion is infinite. If one goes away from that limit, the relevant theories differ from each other accordingly. In our work we have also considered the effect of the RSOC, which was not taken into account in Ref. \cite{ferraz_2023_ConnectionKitaev_AnnalsOfPhysics}.

\section{Presence of RSOC ($\alpha \neq 0$)}
\label{sec:presence-rsoc-alpha}
\begin{figure}
  \centering
\includegraphics[width=0.48\textwidth]{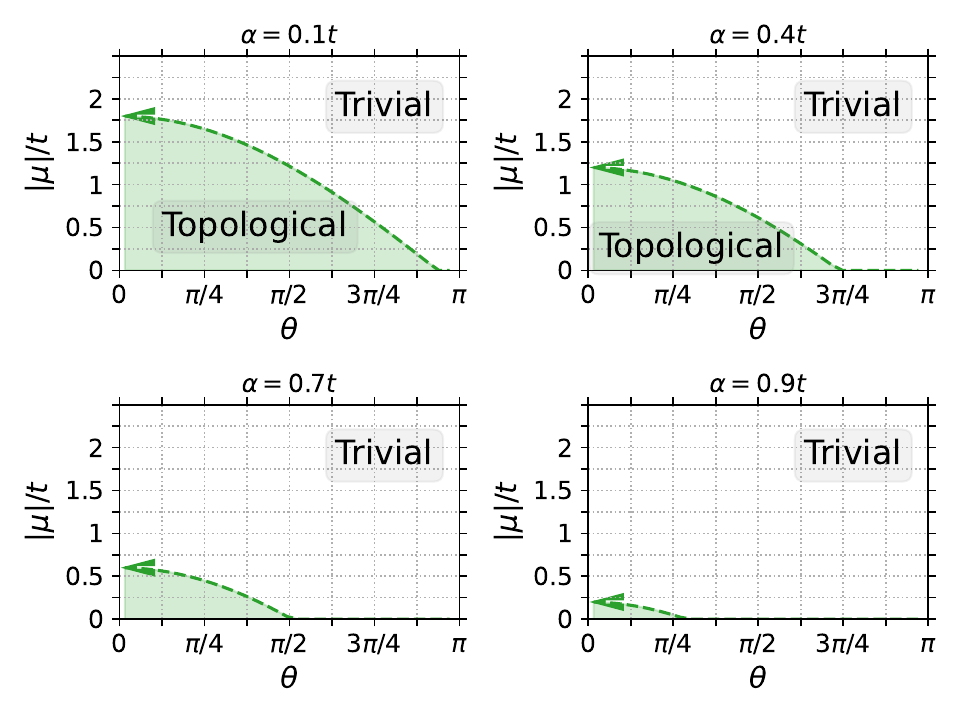}
  \caption{Dependence of trivial and topological phase on $\theta$ for different Rashba interaction $\alpha=0.1t$, $0.2t$, $0.7t$, and $0.9t$. The plotted condition for topological phase is found from Eq. (\ref{eq:condn-rsoc-top-phase}): $\left| \mu \right|/t=2 \cos (\theta/2)- \left| \alpha \right|/t$.}
  \label{fig:alpha-bound}
\end{figure}
When the RSOC is present, i.e. when $(\alpha \neq 0)$, we can observe in the $|\mu|/t$-$\theta$ phase plane the region in which the topological phase manifest is contracted with the increase in the value of $\alpha$. To show this more explicitly we rewrite the effective Hamiltonian, Eq. (\ref{eq:effective-ham-simp-spin-config}), as:
\begin{equation}
  \label{eq:effective-ham-simp-spin-config-rsoc}
  \begin{aligned}
    &H_{\text{eff}}(z, \xi) = && - \sum\limits_{i} \tilde{t}_{i} \: \bar{\xi}_{i} \xi_{i+1}
    - \Delta\sum\limits_{i}\bar{\xi}_{i}\bar{\xi}  _{i+1} \sin \frac{\theta}{2}\\
    & &&\qquad + \text{H.c.} + \mu \sum\limits_{i} \bar{\xi}_{i} \xi_{i},\\
    &\text{where}\\
    & &&\tilde{t}_{i} \equiv \left[ t \cos \left( \frac{\theta}{2} \right) - \alpha \sin \left( \theta_{i} +\frac{\theta}{2} \right) \right]
  \end{aligned}
\end{equation}
is the modified hopping term. Interestingly, it is spatially oscillating through $\sim \alpha\sin \theta_{i}$. Eqs. (\ref{eq:effective-ham-simp-spin-config-rsoc}) and (\ref{eq:effective-ham-kitaev}) are analogous to each other, apart from the modified hopping parameter. Therefore the boundary of the topological phase can be established by the same process --- one can canonically represent the fermionic $\xi_{i}$ operators in terms of the Majorana fermions, and apply the limiting conditions to find the associated topological phase boundary (see. Ref. \cite{kitaev_2001_UnpairedMajorana_Phys-Usp}). The resulting condition for the topological phase is, then, $\left| \mu \right| < \text{Min.}\left(2 \tilde{t}_{i} \right)$. Explicitly we can write:
\begin{equation}
  \label{eq:condn-rsoc-top-phase}
  \left| \mu \right| < 2 \left( t \cos \frac{\theta}{2} - \left|\alpha \right| \right).
\end{equation}
The corresponding values for $\mu$, $t$, and $\theta$ which determine both the topological and trivial phases for different values of $\alpha$ are displayed in Fig. \ref{fig:alpha-bound}. We can note that, the parameter range for the topological phase decreases gradually as we increase the magnitude of $\alpha$. When $\alpha \geq t \cos \left( \theta/2 \right)$ the topological phase is completely absent. Therefore, the RSOC and non alignment of the spin vectors negatively affect the topological phase in a strongly correlated regime. We can conclude that for strongly correlated wire the RSOC does not add new physics to the problem.

\section{Superconducting Condensation energy}
\label{sec:internal-energy}
\begin{figure}
  \centering
\subfloat[]{\label{fig:Int-Ener-Cons-delta-diff-mu-a}\includegraphics[width=.24\textwidth]{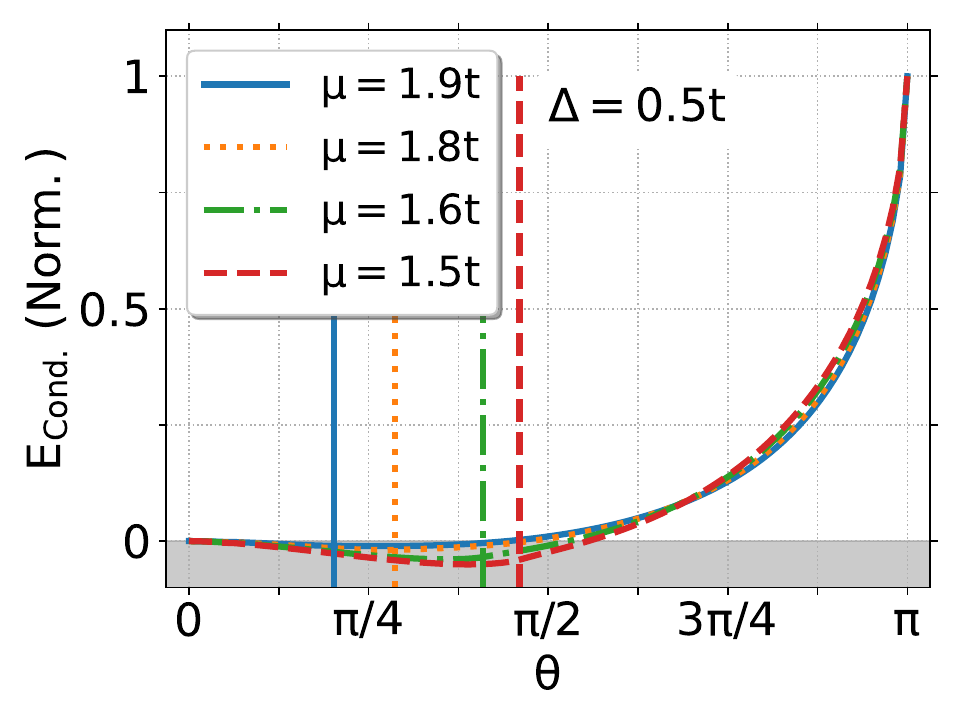}}
  \subfloat[]{\label{fig:Int-Ener-Cons-delta-diff-mu-b}\includegraphics[width=.24\textwidth]{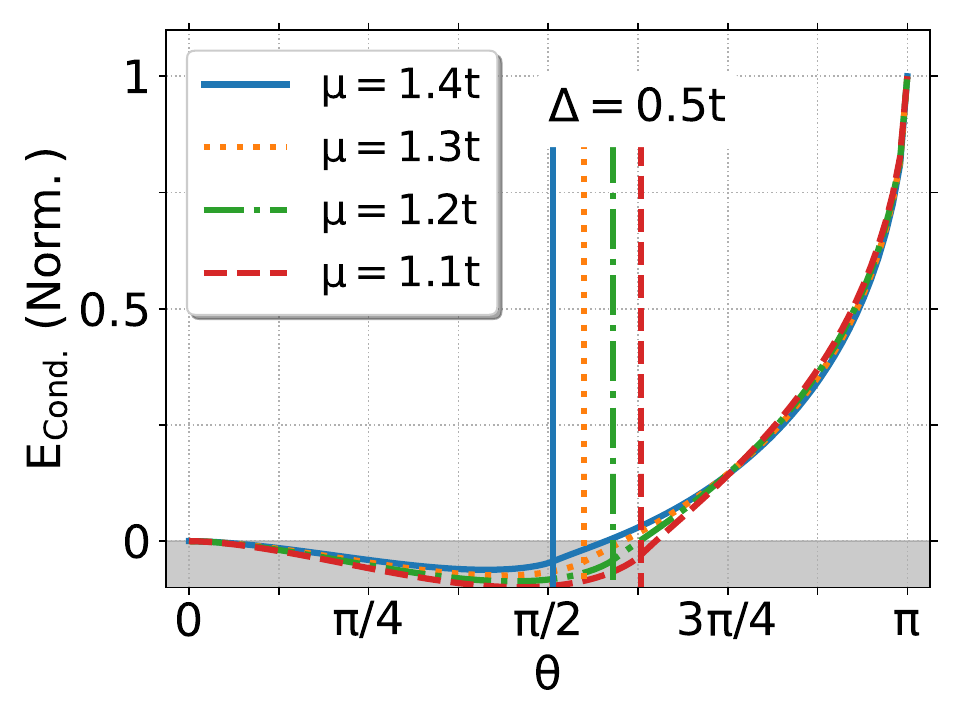}}\\
  \subfloat[]{\label{fig:Int-Ener-Cons-delta-diff-mu-c}\includegraphics[width=.24\textwidth]{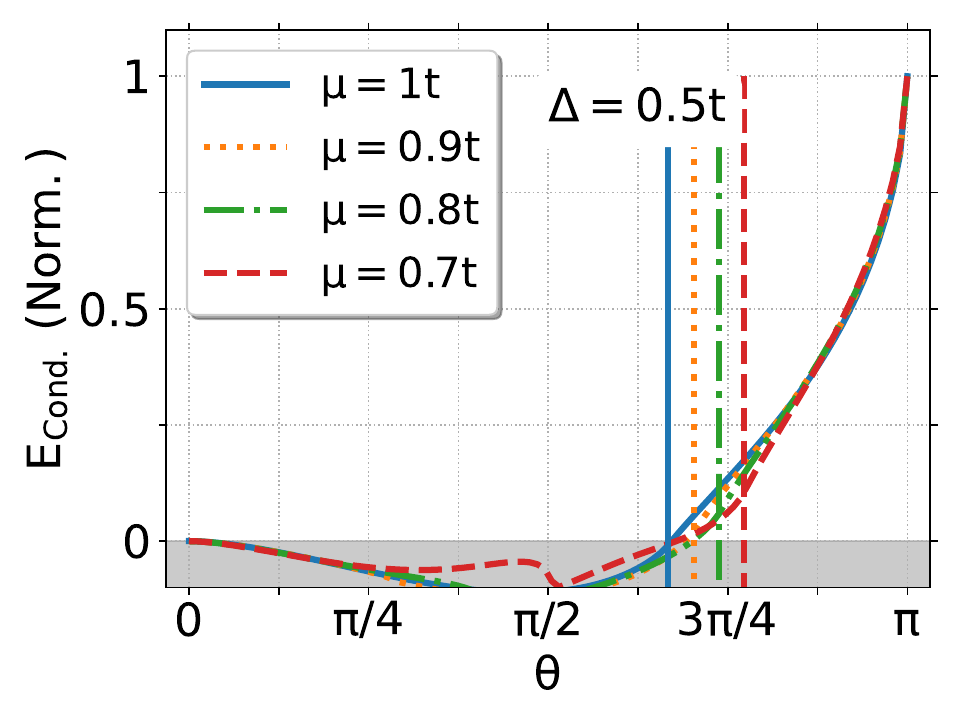}}
  \subfloat[]{\label{fig:Int-Ener-Cons-delta-diff-mu-d}\includegraphics[width=.24\textwidth]{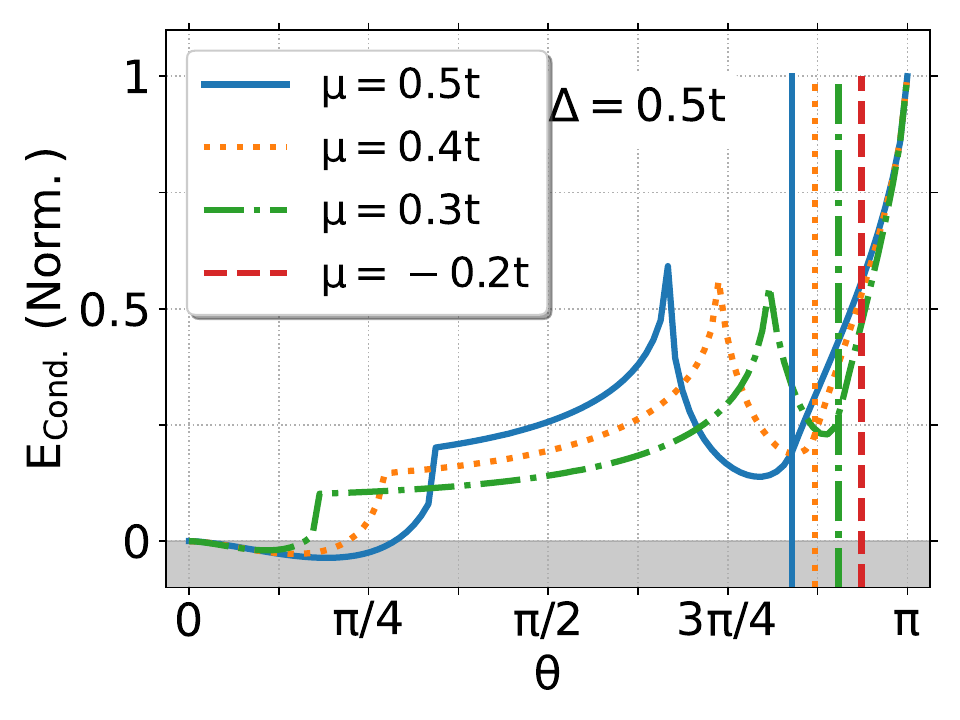}}\\
  \caption{(a,b,c,d) Dependence of the condensation energy ($E_{\text{Cond.}}$) on $\theta$ for constant superconducting gap ($\Delta=0.5t$) as the chemical potential ($2t>\mu>0$) decreases. The $E_{\text{Cond.}}$ is found by numerically integrating Eq. (\ref{eq:Internal-Energy-Def}) over Brillouin zone $k \in \left[ -\pi,\pi \right]$ and later applying Eq. (\ref{eq:conden-energy}). The $E_{\text{Cond.}}$ for different $\mu$, $\Delta$ are normalized. The $\theta_{\text{Topo}}$ found from Eq. (\ref{eq:sufficient-condn-theta}) is plotted as vertical bar. Values of $\theta$ on the right of this bar where maxima occur have topological phase. Hence, if the energy maxima lie on the right of this vertical bar (but $\theta \neq \pi$) then the system thermodynamically resides in the topological phase.}
  \label{fig:Int-Ener-Cons-delta-diff-mu}
\end{figure}
The results found in the previous sections were applicable when the spiral spin texture is present in the system. In Eq. (\ref{eq:energ-disp-eff-Kit-Ham}) it was explicitly shown that the single mode energy depends also on the spin rotation angle $\theta$. However, as mentioned before, the background spin field is dynamic in nature. This means that $\theta$ changes as we change the global system parameters $\mu$ and $\Delta$. These two parameters are essential for the determination of the lowest internal energy ($U_{\text{Internal}}$). The internal energy is found by summing over the energies of all the occupied states:
\begin{equation}
  \label{eq:Internal-Energy-Def}
  \begin{aligned}
    &U_{\text{Internal}} = \sum\limits_{k \in BZ} E(k) f \left[ E(k), \mu \right],\\
    \text{where,}\\
    &f \left[ E(k),\mu \right] \equiv \frac{1}{\mathrm{e}^{\left[ E(k) -\mu\right]/k_{B}T} + 1}.
  \end{aligned}
\end{equation}
$f \left[ E(k), \mu \right]$ is the Fermi-Dirac distribution; $k_{B}=8.6\times 10 ^{-5}$ eV K$^{-1}$ is the Boltzman constant; $T$ is the temperature. In our calculation we take $T \approx 0$ K. The summation in Eq. (\ref{eq:Internal-Energy-Def}) is taken over the whole Brillouin zone $k \in \left[ -\pi, \pi \right]$. Although $U_{\text{Internal}}$ is an important characteristic of the system, for superconducting systems the condensation energy $E_{\text{Cond.}}$ is of importance. Physically it represents the decrease in internal energy in the superconducting states $U_{\text{Internal}}\left( \Delta \neq 0 \right)$ compared to the normal state $U_{\text{Internal}}\left( \Delta = 0 \right)$. It is defined as \cite[see Sec. 3.4.2 of Ref.][]{tinkham_2004_IntroductionSuperconductivity_}:
\begin{equation}
  \label{eq:conden-energy}
  E_{\text{Cond.}} = U_{\text{Internal}}\left( \Delta = 0 \right) - U_{\text{Internal}}\left( \Delta \neq 0 \right).
\end{equation}
Positive $E_{\text{Cond.}}$ means decreases in internal energy in superconducting states compared to non superconducting states, and \emph{vice-versa}. As it is always preferable that the ground state of the system has the lowest possible energy, hence, the system always tries to be in that state where the $E_{\text{Cond.}}$ is highest.

\begin{figure}
  \centering
\subfloat[]{\label{fig:Int-Ener-Cons-mu-diff-delta-a}\includegraphics[width=.24\textwidth]{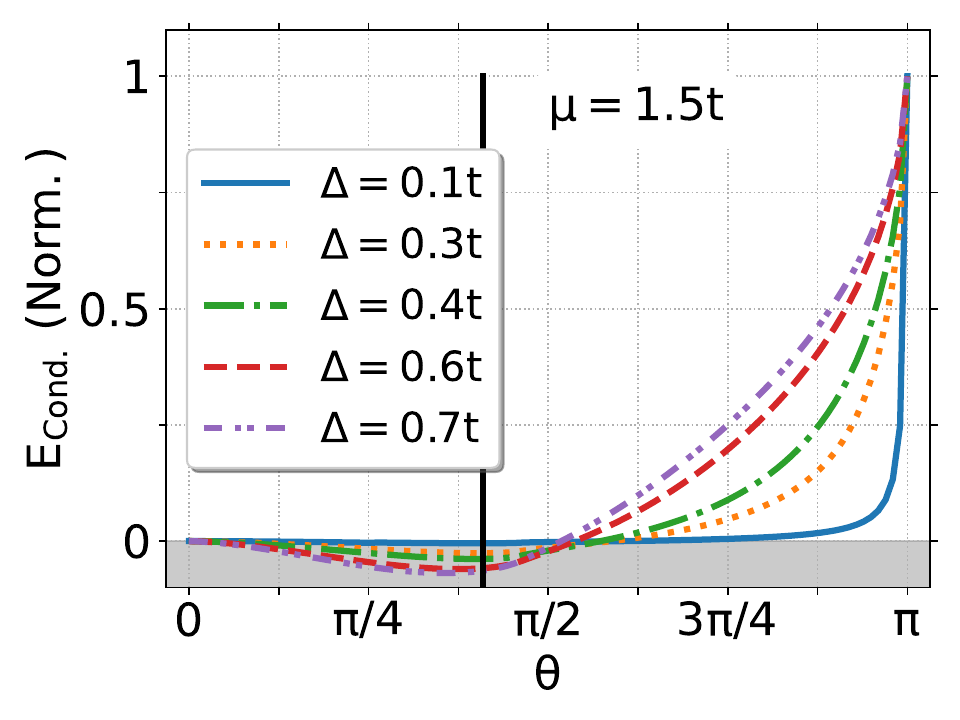}}
  \subfloat[]{\label{fig:Int-Ener-Cons-mu-diff-delta-b}\includegraphics[width=.24\textwidth]{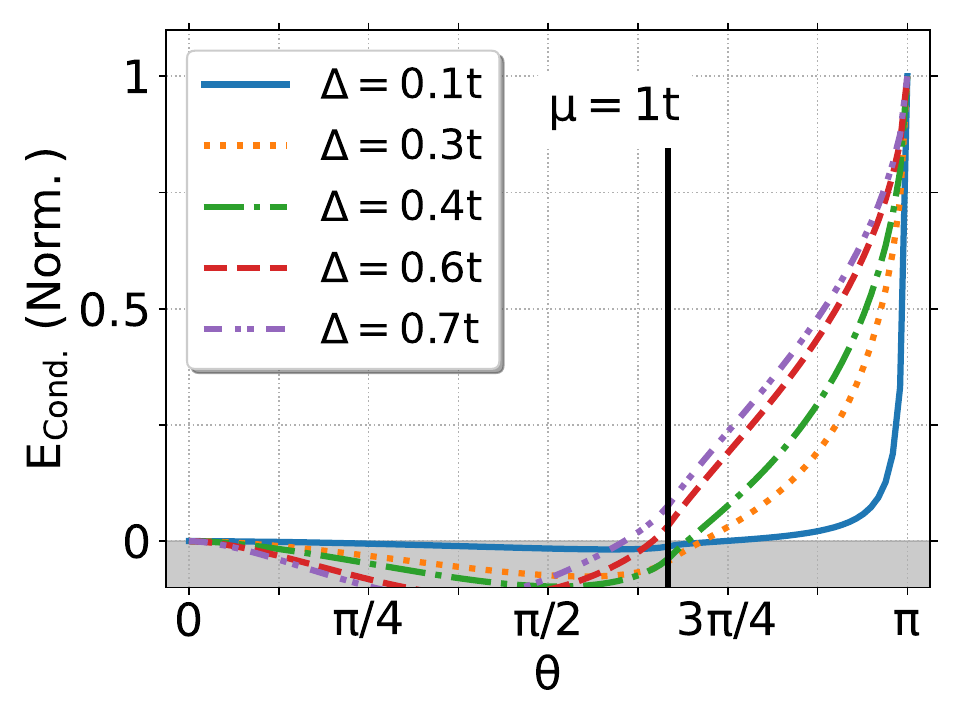}}\\
  \subfloat[]{\label{fig:Int-Ener-Cons-mu-diff-delta-c}\includegraphics[width=.24\textwidth]{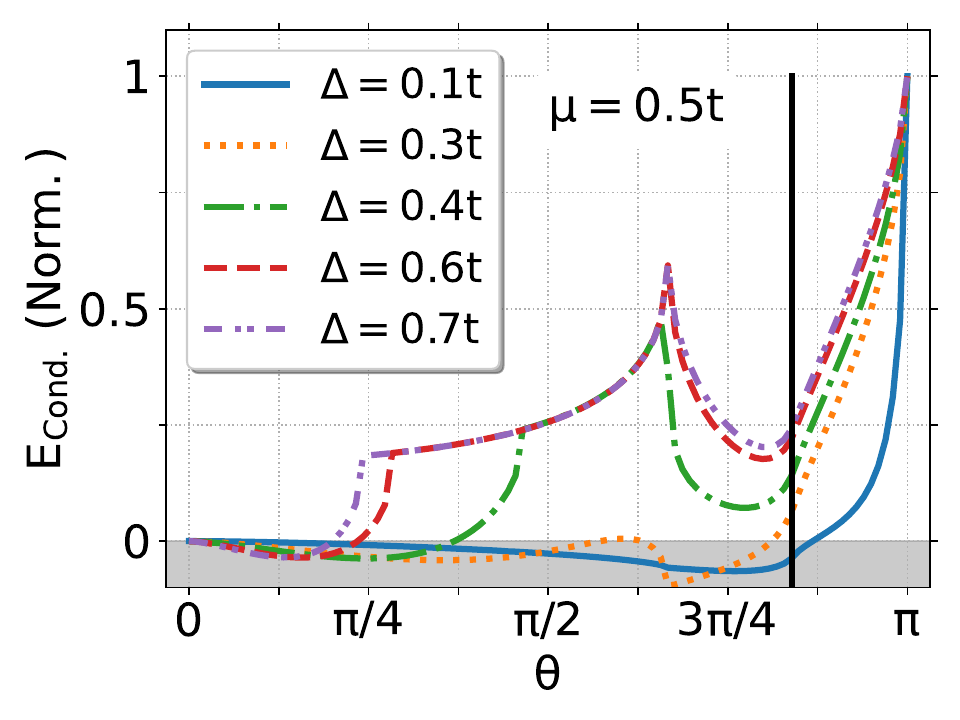}}
  \subfloat[]{\label{fig:Int-Ener-Cons-mu-diff-delta-d}\includegraphics[width=.24\textwidth]{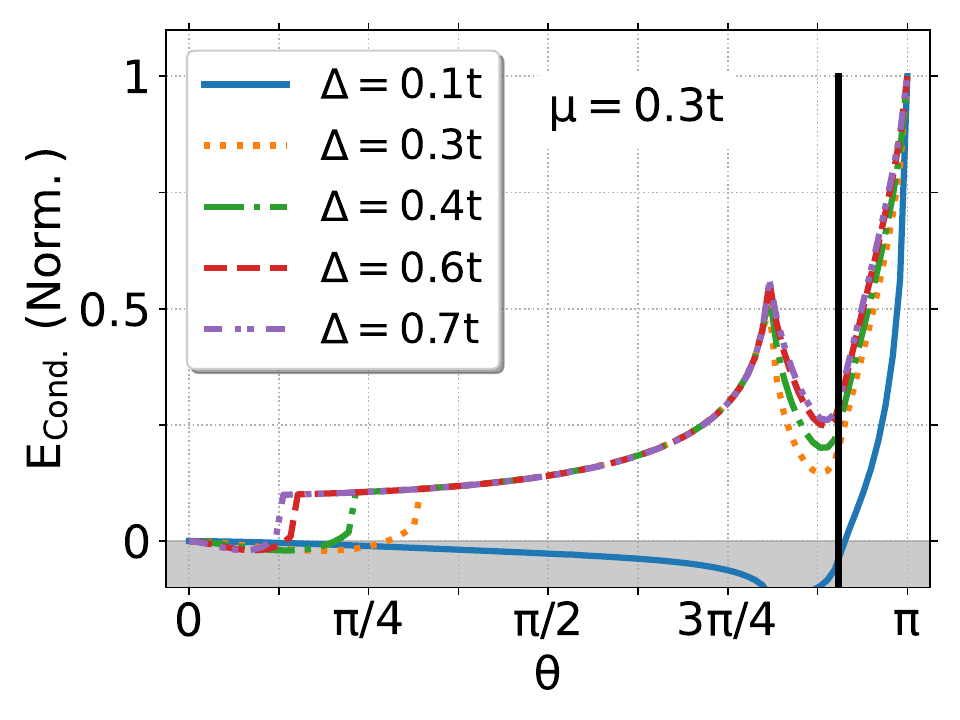}}\\
  \caption{(a,b,c,d) Dependence of the condensation energy ($E_{\text{Cond.}}$) on $\theta$ for constant chemical potential ($\mu$) as the superconducting gap ($\Delta$) increases. The $E_{\text{Cond.}}$ is found by numerically integrating Eq. (\ref{eq:Internal-Energy-Def}) over Brillouin zone $k \in \left[ -\pi,\pi \right]$ and later applying Eq. (\ref{eq:conden-energy}). The $E_{\text{Cond.}}$ for different $\mu$, $\Delta$ are normalized. The $\theta_{\text{Topo}}$ found from Eq. (\ref{eq:sufficient-condn-theta}) is plotted as vertical bar. Values of $\theta$ on the right of this bar at which the maxima occur have topological phase. Hence, if the energy maxima lie on the right of this vertical bar (but $\theta \neq \pi$) then the system thermodynamically resides in the topological phase.}
  \label{fig:Int-Ener-Cons-mu-diff-delta}
\end{figure}

In Fig. \ref{fig:Int-Ener-Cons-delta-diff-mu} we show the dependence of the condensation energy on $\theta$ for constant $\Delta=0.5t$, and for different $\mu$ values. The energies are normalized in these figures. As positive $E_{\text{Cond.}}$ is preferable we search for local maxima in these plots. It should be remembered that the system has topological properties only when $\theta \neq 0, \pi$; this is a necessary but not sufficient condition. The second condition is:
\begin{equation}
  \label{eq:sufficient-condn-theta}
  \theta > \theta_{\text{Topo}} \equiv 2 \: \arccos \left|\frac{\mu}{2 t} \right|.
\end{equation}
This gives the lower boundary $\theta_{\text{Topo}}$ for a topological phase; $\theta_{\text{Topo}}$ is a function of $\mu$. In Fig. \ref{fig:Int-Ener-Cons-delta-diff-mu} the values of $\theta_{\text{Topo}}$ obtained from Eq. (\ref{eq:sufficient-condn-theta}) are displayed as vertical lines. Therefore if a global or local maximum lies on the right of the vertical bar and does not occurs at $\theta = \pi$, then the system will thermodynamically acquire a topological phase. In Fig. \ref{fig:Int-Ener-Cons-mu-diff-delta} we show the condensation energy dependence on $\theta$ for several values of $\mu$ and $\Delta$. The values of $\theta$ found from Eq. (\ref{eq:sufficient-condn-theta}) is shown as the vertical line.

\begin{figure}[tbh]
  \centering
   \includegraphics[width=.48\textwidth]{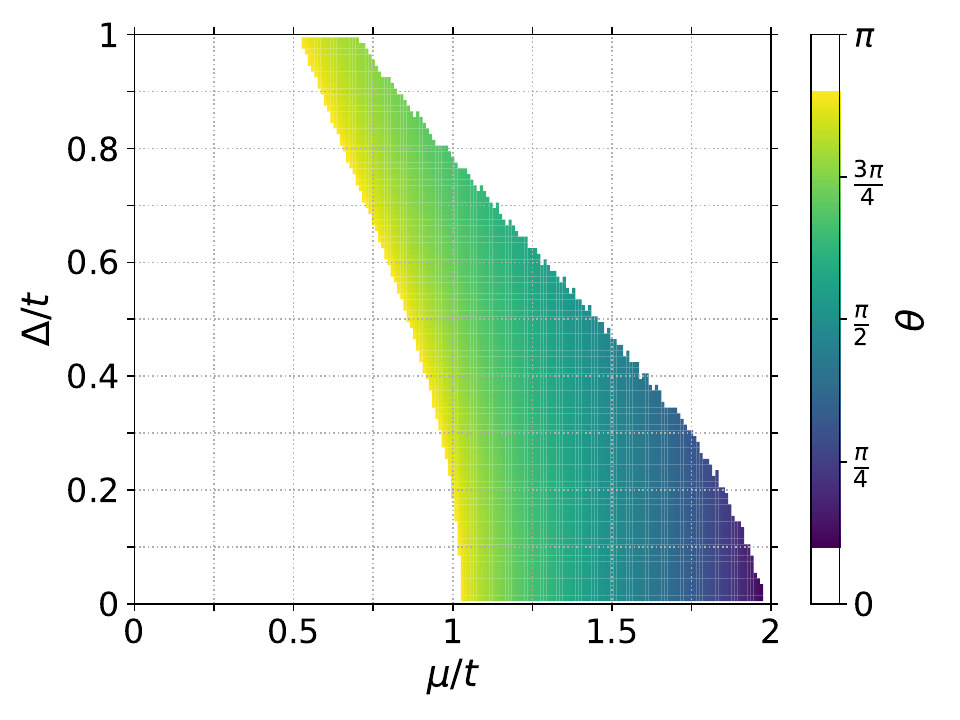}
   \caption{Phase diagram showing the values of $\mu/t$ and $\Delta/t$ for which the topological phase occurs thermodynamically. Topological phase appears for the shaded regions, however, it is absent for unshaded regions. The color coding of the shaded region represent the values of $\theta$.} 
  \label{fig:mu-delta-phase-diag}
\end{figure}
As was mentioned before the following two conditions are sufficient and necessary for the emergence of the topological phase thermodynamically: (i) $\theta \in 0, \pi$, (ii) $\theta > \theta_{\text{Topo}}$; $\theta$ being the angle where the maxima of $E_{\text{Cond.}}$ occurs. In Fig. \ref{fig:mu-delta-phase-diag} we plot the values of $\mu$ and $\Delta$ which satisfy the aforementioned conditions thermodynamically. The shaded regions represents the topological phase, while the unshaded regions represent the trivial no topological phase. The color coding in the figure represents the value of $\theta$ where maximum energy occurs thermodynamically.

\begin{figure}[tbh]
  \centering
   \includegraphics[width=.48\textwidth]{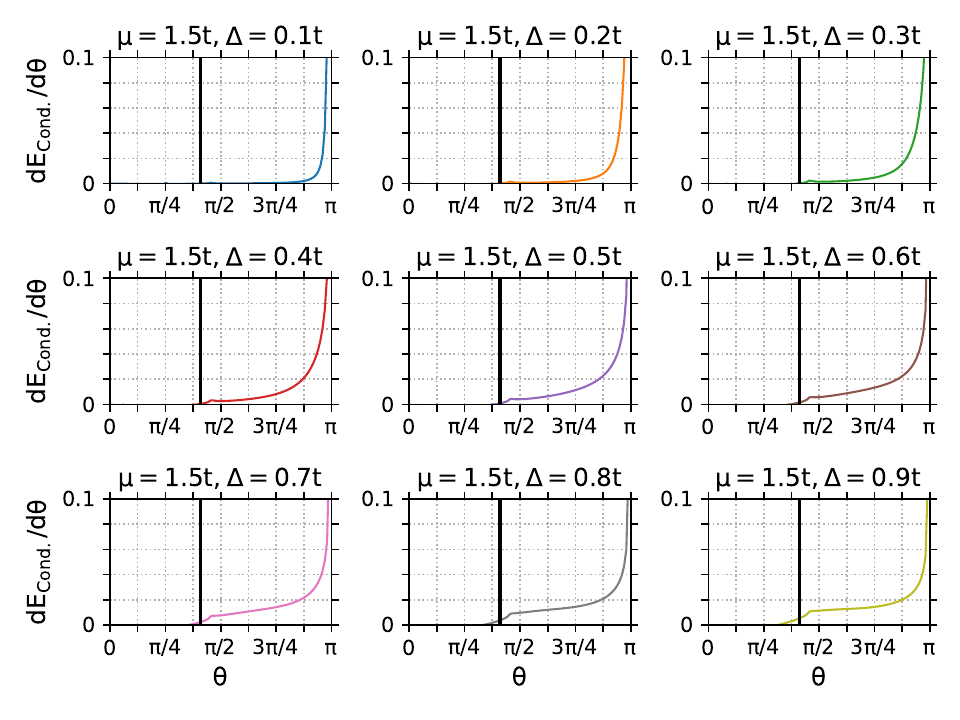}
   \caption{Differentiation of condensation energy \emph{w.r.t} angle $\theta$ $\left( dE_{\text{Cond.}}/d \theta \right)$ for $\mu=1.5 t$ and different $\Delta$. The value of $\theta$ where $\left( dE_{\text{Cond.}}/d \theta \right)=0$ local maximum or plateau occurs. The black vertical bar is the $\theta_{\text{Topo.}}$ for $\mu=1.5t$.}
  \label{fig:dE-dtheta-dep}
\end{figure}

In addition to Fig. \ref{fig:mu-delta-phase-diag}, in Fig. \ref{fig:dE-dtheta-dep} we also show the first order differentiation of the condensation energy $dE_{\text{Cond.}}/d\theta$ for constant $\mu=1.5t$ and different $0.1 \leq \Delta \leq 0.9$. From the plot of the dependence of $E_{\text{Cond.}}$ on $\theta$ for $\mu=1.5t$ (Fig. \ref{fig:Int-Ener-Cons-mu-diff-delta}a), we see that 
all the curves monotonically decrease (except $\Delta=0.1t$) before increasing. In the portion of the plot where the plot increases if the $dE_{\text{Cond.}}/d\theta=0$ then either a local maxima of plateau occurs. If these special points lies on the right of the $\theta_{\text{Topo.}}$ then the topological phase occurs thermodynamically. One can observe that for $\Delta<0.5t$ the $dE_{\text{Cond.}}/d\theta=0$ is satisfied for $\theta>\theta_{\text{Topo.}}$.

The main physical conclusion from this section is that by modifying the chemical potential we can drive the system into and out of the topological phase thermodynamically. Although previously proposed heterostructures \cite{lutchyn_2018_MajoranaZero_NatRevMater} used the same broad idea, in this work no external magnetic field is needed to produce such an effect.

\section{Proposal for experimental heterostructure}
\label{sec:prop-exper-heter}
\begin{figure}
  \centering
  \subfloat[]{\label{fig:device-schematics-a}\includegraphics[width=.35\textwidth]{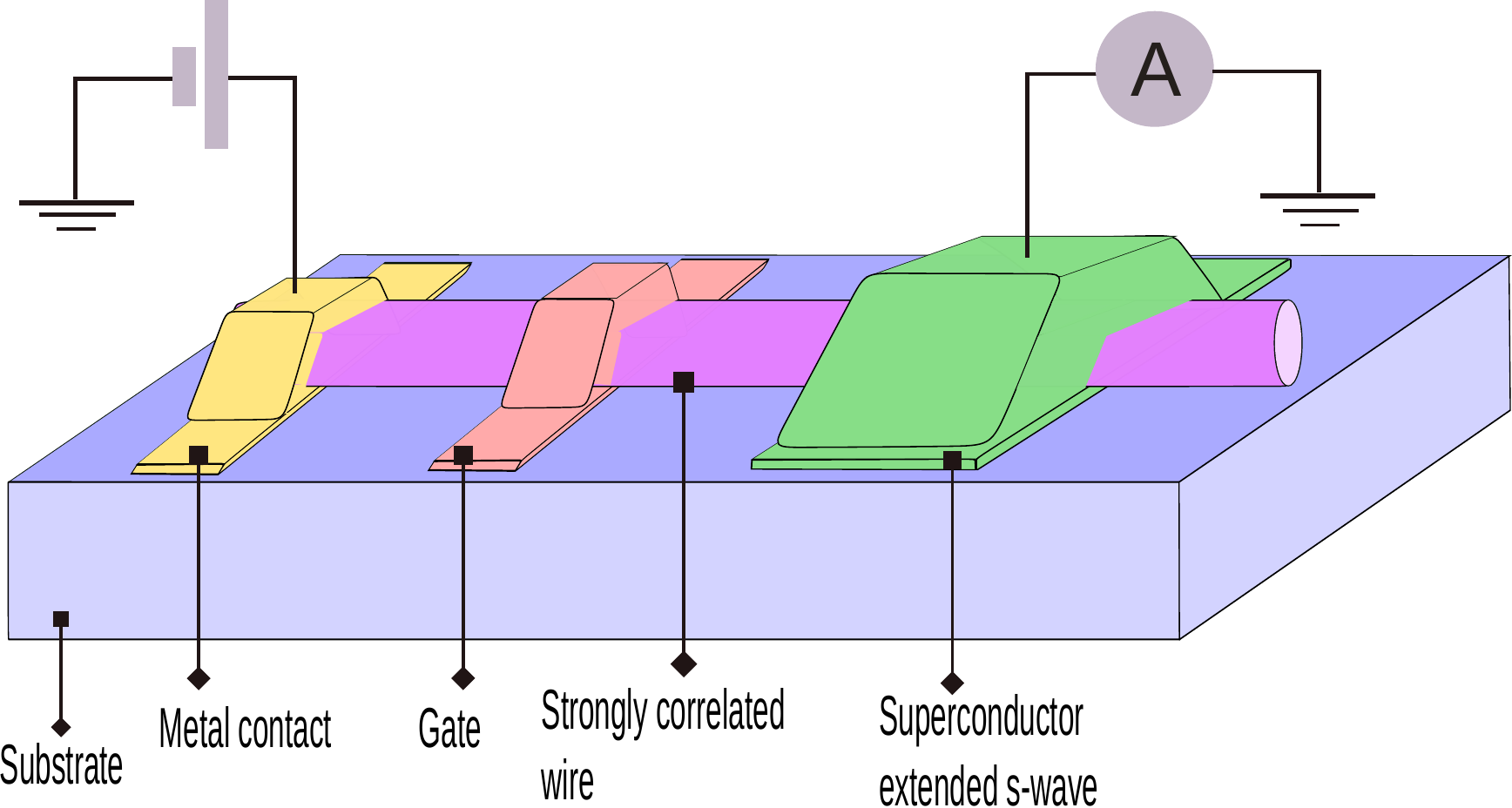}}\\
  \subfloat[]{\label{fig:device-schematics-b}\includegraphics[width=.35\textwidth]{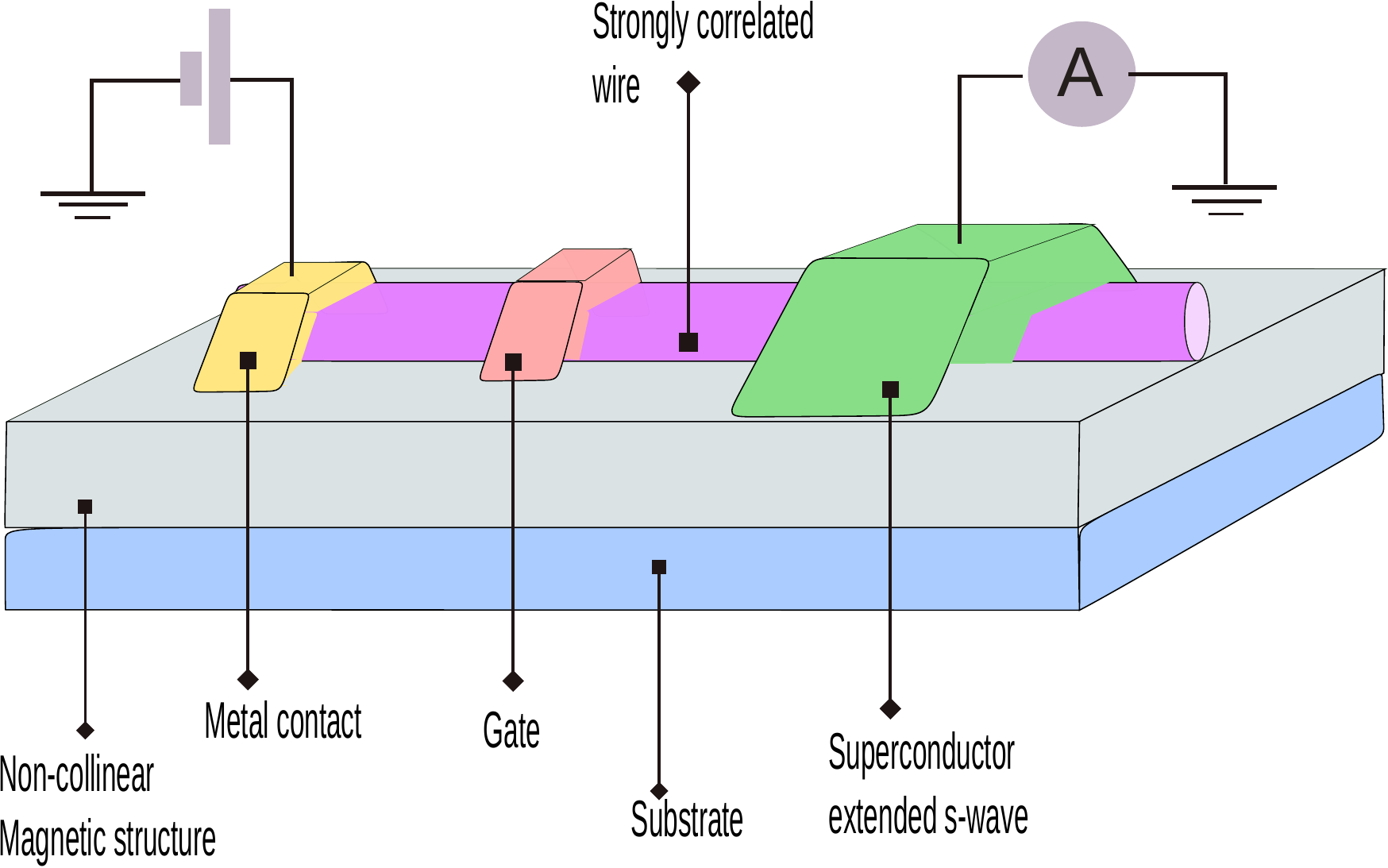}}
   \caption{Device schematics: (a) Tunneling experiment probing the presence of the Majorana fermions due to dynamic spin field of the 1D nanowire. (b) Tunneling experiment due to induced non-collinear spin texture through bottom layer. One can use the Van der Waals magnets as the non-collinear magnetic layer.}
  \label{fig:device-schematics}
\end{figure}
Usually tunneling experiments are employed to detect the Majorana fermions \cite{mourik_2012_SignaturesMajorana_Science}. The proposed experimental setups for detecting Majorana fermions in strongly correlated wires are shown in Fig. \ref{fig:device-schematics}. A zero bias peak in the conductance plot is usually interpreted as the possible signature of the MZM. However, these experiments are not at all conclusive, because Andreev bound states also have the same zero bias peak signature \cite{prada_2020_AndreevMajorana_NatRevPhys}. Despite these difficulties, recently a number of experiments have been performed to detect the MZM \cite{lutchyn_2018_MajoranaZero_NatRevMater,cao_2023_RecentProgress_SciChinaPhysMechAstron,nadj-perge_2014_ObservationMajorana_Science}. We propose analogous experiments to detect MZM.

In Fig. \ref{fig:device-schematics-a} a strongly correlated wire is placed on top of an insulating substrate. The tunneling electrons enters into the wire out through the metallic contacts. Furthermore, they exit the wire out through the superconducting contact, and flow directly into the current measuring device. The chemical potential is controlled through the gate contacts. For nanowires either heavy fermion \cite{morosan_2012_StronglyCorrelated_AdvMater,paschen_2021_QuantumPhases_NatRevPhys,carvalho_2022_PossibleRoutes_JPhys:ConfSer,moura_2016_DimensionalityTuning_SciRep,rosa_2014_ExploringEffects_SolidStateCommunications,yang_2022_OrderedPtFeIr_AngewChemIntEd} or transition metal \cite{nautiyal_2004_NanowiresSpintronics_PhysRevB,delin_2003_MagneticPhenomena_PhysRevB,lim_2021_WaferScaleGrowth_NanoLett,luo_2009_Magnetism3dTransition_JPhysChemC,lim_2022_NanowiretoNanoribbonConversion_ACSApplNanoMater,li_2022_TunableStrong_AdvancedMaterials} compounds can be used. Especially nanowires from transition metal dichalcogenides compounds are promising candidates \cite{li_2022_TunableStrong_AdvancedMaterials,lim_2021_WaferScaleGrowth_NanoLett,lim_2022_NanowiretoNanoribbonConversion_ACSApplNanoMater}. For superconductors one can use the iron-based materials, since they have extended \emph{s}-wave order parameter \cite{stewart_2017_UnconventionalSuperconductivity_AdvPhys,stewart_2011_SuperconductivityIron_RevModPhys,bang_2017_SuperconductingProperties_JPhys:CondensMatter,biswal_2021_RecentReview_MaterialsToday:Proceedings}. Because the spiral spin field is dynamically generated in strongly correlated 1D nanowires, we expect to see a MZM signature in the tunneling experiments, without the need of an external applied rotating magnetic field to generate spiral spin fields.

However one can also perform experiments by externally inducing a non-collinear spin textures in the wire. The schematics for this corresponding experiment is shown in Fig. \ref{fig:device-schematics-b}. Here, the only difference from the previous arrangement (Fig. \ref{fig:device-schematics-a}) is that, we place the nanowire on a magnetic layer with non-collinear magnetic structures. The magnetic structure of the base magnetic material is induced into the nanowire, which is a necessary condition for the MZM to occur. One can also use the Van der Waals magnets as magnetic layer \cite{wang_2022_MagneticGenome_ACSNano,elahi_2022_ReviewTwodimensional_ComputationalMaterialsScience}. They may host different magnetic structures, e.g. Skyrmions, spirals \cite{klein_2022_ControlStructure_NatCommun,yang_2023_MoireMagnetic_NatComputSci,meijer_2020_ChiralSpin_NanoLett,casas_2023_CoexistenceMerons_AdvancedMaterials,wu_2022_VanWaals_AdvMater,hejazi_2020_NoncollinearPhases_ProcNatlAcadSci,wang_2020_StackingDomain_PhysRevLett,wu_2020_NeeltypeSkyrmion_NatCommun,tong_2019_MagneticProximity_PhysRevAppl,yang_2020_CreationSkyrmions_SciAdv,tong_2018_SkyrmionsMoire_NanoLett,burch_2018_MagnetismTwodimensional_Nature}.

Finally, there is yet a third way to perform the required experiment. Instead of the nanowire one can use a single wall carbon nanotube, which has the required strong correlation \cite{izumida_2009_SpinOrbit_JPhysSocJpn,jeong_2009_CurvatureenhancedSpinorbit_PhysRevB,li_2019_SingleWalledCarbon_}. Interestingly, in these nanotubes RSOC can be induced and controlled through electric field \cite{min_2006_IntrinsicRashba_PhysRevB,klinovaja_2011_HelicalModes_PhysRevLett,novikov_2006_EnergyAnomaly_PhysRevLett}. The schematic of the device is analogous to the one displayed in Fig. \ref{fig:device-schematics-a}; instead of the nanowire, a single wall carbon tube is placed over the metal contact, the gate and the superconducting contact. One should remember, in this case that the applied electric field also controls the RSOC.

The aforementioned experimental setup are known as local detection of MZM, however, there also exist several non-local detection of the MZM \cite{zhang_2019_NextSteps_NatCommun}. In one of the scheme, basically one adds two extra gates to the two ends of the nanowires. The gate near the tunnel gate is denoted as local gate, and the other is denoted as non-local gate. The local gate measures the local density of states ($dI/dV$). When the non-local gate is tuned the MZM on the other side move towards the MZM at the local gate, and finally two MZMs merges with each other to create an interference pattern. It can be detected by the transformation of the $dI/dV$ from single peak at $V=0$ to the double peak on two sides of $V=0$. There also exists the three terminal circuits where two tunnel gates are placed at two side of the superconducting induced wire. As these tunnel gate measures the local density of states on the two sides, one can use measure the $dI/dV$ on the two sides of the wire. It the MZM is present at two ends of the nanowire, the zero bias peak corresponding to these two MZM should be correlated. Apart from electric measurement there also exist the caloric function measurement of the MZM \cite{shustin_2022_FeaturesPhysical_JExpTheorPhys,aksenov_2020_StrongCoulomb_PhysRevB}.

  For application of our model to real systems following energy scales of different parameters should be satisfied. If $a$ is the lattice constant and $m$ is the electric effective mass, then the hopping strength is $t=\hbar^{2}/ma^{2}$. We assume that the RSOC in the wire is comparable to the values in two dimensional quantum wells having heavy elements; RSOC in these materials in continuous limit is $\alpha_{cont.} \sim 10^{4}$---$10^{5}$ m/s. However as we are considering the lattice model, hence instead of continuous limit RSOC ($\alpha_{cont.}$), we need to use the lattice RSOC ($\alpha_{Lat.}$). The $\alpha_{cont.}$ is related to the lattice limit as: $\alpha_{Lat.}=\frac{\hbar}{a}\alpha_{cont.}$. In semiconductor with heavy elements the electron effective mass is of order $m \sim 0.05 m_{e}$. Putting all the values together we will get $m\alpha_{cont.}^{2}=\alpha_{Lat.}^{2}/t \sim 1$ K. For a reasonable proximity effect $\Delta \sim $ 1---10 K. Hence the relevant hierarchy of energy will be: $U \gg t > \alpha_{Lat.}>\Delta$. In the absence of RSOC one can put $\alpha_{Lat.}=0$.

We should make a small comment about the extended $s$-wave iron based superconductors. One usually expects that the order parameter in this case will have nodes. It means the close in the superconducting gap, which will destroy the topological phase. However, recent experimental works, although not fully convincing, have shown otherwise \cite[see Sec. VIII of Ref. ][]{qin_2022_TopologicalSuperconductivity_PhysRevX}. It was argued that spin-orbit coupling and inter-band pairing can give rise to the nodeless superconducting order parameter.

\section{Conclusion}
\label{sec:conclusion}
In this work, it was shown that the Majorana zero modes in nanowire-superconductor heterostructures can emerge without application of an external magnetic field; the only condition is that, the electrons in the nanowire should be required to be strongly correlated. Within a mean-field treatment of Eq. (\ref{eq:ham-1d-wire}) that ignores the no double occupancy constraint, the Hubbard operators are replaced with the conventional electron operators, $X^{\sigma 0}\to c^{\dagger}_{\sigma}$, $X^{0\sigma}\to c_{\sigma}$, $X^{0 0}\to \sum\limits_{\sigma}c_{\sigma}^{\dagger}c_{\sigma}$, \emph{etc}. As a result, the Hamiltonian reduces to an exactly solvable one that produces the doubly degenerate band structure. In order to effectively generate the spinless fermions one needs to separate the bands by adding into consideration a large external magnetic field. This however may totally destroy the superconductivity. To avoid such a scenario, we apply the mean-field treatment only after the no double occupancy constraint has rigorously been imposed. Within our approach, there is no electron double occupancy since the spinless fermion field $\xi_i^2\equiv 0$. The spinless fermion excitations emerge in the absence of an external magnetic field. This is the reason we don't need magnetic field when strong \emph{e-e} correlation is present.

The physical system shown in Fig. \ref{fig:schematic} was investigated for a spiral spin field, both in the presence and absence of the Rashba spin orbit coupling (RSOC). It was found that, apart from the chemical potential ($\mu$), an extra parameter $\theta$ --- the angle between spin projection of the neighboring sites on $xy$ plane --- enters into the Hamiltonian as a tuning parameter for allowing the system to go into and out of the topological phase. This can be easily observed by comparing the single mode energy as described by, Eq. (\ref{eq:energ-disp-eff-Kit-Ham}), with the same single mode energy displayed by the Kitaev's toy model \cite{alicea_2012_NewDirections_RepProgPhys,kitaev_2001_UnpairedMajorana_Phys-Usp}. The increase in $\theta$ decreases the allowed values of $\mu$ and $t$ for which the system enters into the topological phase, as shown in Fig. \ref{fig:kitaev-phase-diag}. We investigated the $\theta$ dependence of the internal energy, and we found out that only for limited values of $\mu$ and $\Delta$ the system thermodynamically enters into the topological phase, as shown in Fig. \ref{fig:Int-Ener-Cons-delta-diff-mu} and \ref{fig:Int-Ener-Cons-mu-diff-delta}. The corresponding phase diagram is shown in Fig. \ref{fig:mu-delta-phase-diag}. We also found out that RSOC is not a necessary ingredient for the appearance of the topological phase. This result is different from the usual semiconductor-superconductor heterostructure used for the realisation of Majorana fermions, where RSOC, as well as, an external magnetic field, are the necessary ingredients. In fact, in our case with increase in intensity of RSOC the area in the parameter phase space ($\mu/t - \theta$) for occurrence topological phase decreases, as shown in Fig. \ref{fig:alpha-bound}. Finally, in Sec. \ref{sec:prop-exper-heter} we proposed possible ways to experimentally realize the proposed system.

\section{Acknowldgements}
K.K.K wish to thank P. A. Maksimov for valuable discussion. K.K.K acknowledges the financial support from the JINR grant for young scientists and specialists, the Foundation for the Advancement of Theoretical Physics and Mathematics ”Basis” for grant \# 23-1-4-63-1. One of us, A.F., acknowledges financial support from the MEC, CNPq (Brazil) and from the Simons Foundation (USA).

\end{document}